%% file: main.tex
\documentclass[aps,prd,reprint,nofootinbib,superscriptaddress]{revtex4-2}
\usepackage{graphicx}
\usepackage{dcolumn}
\usepackage{bm}
\usepackage[table,x11names,dvipsnames,table]{xcolor}
\usepackage{siunitx}
\usepackage{makecell}
\usepackage{amsmath}
\usepackage{etoolbox}
\usepackage{booktabs}
\usepackage{array}
\usepackage{tabularx}

\usepackage{bigstrut}
\usepackage{multirow}
\usepackage{float}

\usepackage{hyperref}

\newcommand{\technote}[1]{\ifcsdef{istechnote}{#1}{}}

\usepackage{listings}
\begin{document}

\preprint{APS/123-QED}

\title{\technote{Technical Note: }Enhanced Ionization Charge Identification in the Short-Baseline Neutrino Program Neutrino Detectors with Deep Neural Networks}


\input{authors}

\collaboration{ICARUS and SBND collaborations for the SBN program}

\date{\today}

\begin{abstract}
\input{abstract}
\end{abstract}

\maketitle


\section{Introduction}
\input{Introduction}

\section{Signal Filtering and DNN ROI Input Pre-Processing}
\label{sec:pre-process}
\input{pre-process}

\section{Training Samples}
\label{sec:samples}
\input{SampleGen}

\section{Network Optimization and Training}
\input{Network}

\section{Results}
\input{Results}

\section{Conclusion}
\input{Conclusion}

\begin{acknowledgments}
\input{acknowledgements}
\end{acknowledgments}

\bibliography{cite}

\technote{
\clearpage
\onecolumngrid
\appendix
\input{Software}
\input{ICARUSFilterOptimizaton}
\clearpage
\input{NetworkOptimization}
}

\end{document}

%% file: authors.tex
\newcommand{\ANL}{Argonne National Laboratory, Lemont, IL 60439, USA}
\newcommand{\Bern}{Universit\"{a}t Bern, Bern CH-3012, Switzerland}
\newcommand{\BNL}{Brookhaven National Laboratory, Upton, NY 11973, USA}
\newcommand{\UCSB}{University of California, Santa Barbara CA, 93106, USA}
\newcommand{\Campinas}{Universidade Estadual de Campinas, Campinas, SP 13083-970, Brazil}
\newcommand{\CTI}{Center for Information Technology Renato Archer, Campinas, SP 13069-901, Brazil}
\newcommand{\Chicago}{University of Chicago, Chicago, IL 60637, USA}
\newcommand{\CIEMAT}{CIEMAT, Centro de Investigaciones Energ\'{e}ticas, Medioambientales y Tecnol\'{o}gicas, Madrid E-28040, Spain}
\newcommand{\CSU}{Colorado State University, Fort Collins, CO 80523, USA}
\newcommand{\Columbia}{Columbia University, New York, NY 10027, USA}
\newcommand{\Edinburgh}{University of Edinburgh, Edinburgh EH9 3FD, United Kingdom}
\newcommand{\ABC}{Universidade Federal do ABC, Santo Andr\'{e}, SP 09210-580, Brazil}
\newcommand{\Alfenas}{Universidade Federal de Alfenas, Po\c{c}os de Caldas, MG 37715-400, Brazil}
\newcommand{\FNAL}{Fermi National Accelerator Laboratory, Batavia, IL 60510, USA}
\newcommand{\Florida}{University of Florida, Gainesville, FL 32611, USA}
\newcommand{\Granada}{Universidad de Granada, Granada E-18071, Spain}
\newcommand{\IIT}{Illinois Institute of Technology, Chicago, IL 60616, USA}
\newcommand{\Imperial}{Imperial College London, London SW7 2AZ, United Kingdom}
\newcommand{\IIS}{Centre for High Energy Physics, Indian Institute of Science, Bangalore 560012, India}
\newcommand{\SaoJose}{Instituto Tecnológico de Aeronáutica, São José dos Campos, SP 12228-900, Brazil}
\newcommand{\Kansas}{University of Kansas, Lawrence, KS 66045, USA}
\newcommand{\Lancaster}{Lancaster University, Lancaster LA1 4YW, United Kingdom}
\newcommand{\Liverpool}{University of Liverpool, Liverpool L69 7ZE, United Kingdom}
\newcommand{\LANL}{Los Alamos National Laboratory, Los Alamos, NM 87545, USA}
\newcommand{\LSU}{Louisiana State University, Baton Rouge, LA 70803, USA}
\newcommand{\Manchester}{University of Manchester, Manchester M13 9PL, United Kingdom}
\newcommand{\Michigan}{University of Michigan, Ann Arbor, MI 48109, USA}
\newcommand{\Minnesota}{University of Minnesota, Minneapolis, MN 55455, USA}
\newcommand{\Holyoke}{Mount Holyoke College, South Hadley, MA 01075, USA}
\newcommand{\NotreDame}{University of Notre Dame, Notre Dame, IN 46556 USA}
\newcommand{\Oxford}{University of Oxford, Oxford OX1 3RH, United Kingdom}
\newcommand{\Penn}{University of Pennsylvania, Philadelphia, PA 19104, USA}
\newcommand{\Palermo}{Universit\`{a} degli Studi di Palermo, Dipartimento di Fisica e Chimica, I-90123 Palermo, Italy}
\newcommand{\QueenMary}{Queen Mary University of London, London E1 4NS, United Kingdom}
\newcommand{\Rutgers}{Rutgers University, Piscataway, NJ, 08854, USA}
\newcommand{\Sheffield}{University of Sheffield, School of Mathematical and Physical Sciences, Sheffield S3 7RH, United Kingdom}
\newcommand{\Sussex}{University of Sussex, Brighton BN1 9RH, United Kingdom}
\newcommand{\Syracuse}{Syracuse University, Syracuse, NY 13244, USA}
\newcommand{\UTK}{University of Tennessee at Knoxville, TN 37996, USA}
\newcommand{\TAMU}{Texas A\&M University, College Station, TX 77843, USA}
\newcommand{\UTA}{University of Texas at Arlington, TX 76019, USA}
\newcommand{\Tufts}{Tufts University, Medford, MA, 02155, USA}
\newcommand{\UCL}{University College London, London WC1E 6BT, United Kingdom}
\newcommand{\VirginiaTech}{Center for Neutrino Physics, Virginia Tech, Blacksburg, VA 24060, USA}
\newcommand{\Warwick}{University of Warwick, Coventry CV4 7AL, United Kingdom}
\newcommand{\CBPFCENTROBRASILEIRO}{CBPF, Centro Brasileiro de Pesquisas Fisicas, Rio de Janeiro, Brazil}
\newcommand{\CERNEUROPEANORGANIZA}{CERN, European Organization for Nuclear Research 1211 Gen\`eve 23, Switzerland, CERN}
\newcommand{\CENTRODEINVESTIGACIO}{Centro de Investigacion y de Estudios Avanzados del IPN (Cinvestav), Mexico City}
\newcommand{\UNIVERSITYOFHOUSTONH}{University of Houston, Houston, TX 77204, USA}
\newcommand{\INDIANINSTITUTEOFSCI}{Indian Institute of Science, Bengaluru, India}
\newcommand{\INFNSEZIONEDIBOLOGNA}{INFN Sezione di Bologna and University, Bologna, Italy}
\newcommand{\INFNSEZIONEDICATANIA}{INFN Sezione di Catania and University, Catania, Italy}
\newcommand{\INFNSEZIONEDIGENOVAA}{INFN Sezione di Genova and University, Genova, Italy}
\newcommand{\INSTITUTODINEURO}{Istituto di Neuroscienze, CNR, Padova, Italy}
\newcommand{\INFNGSSILAQUILAITALY}{INFN GSSI, L’Aquila, Italy}
\newcommand{\INFNLNGSASSERGIITALY}{INFN LNGS, Assergi,  Italy}
\newcommand{\INFNLNSCATANIAITALY}{INFN LNS, Catania, Italy}
\newcommand{\INFNSEZIONEDIMILANOM}{INFN Sezione di Milano, Milano, Italy}
\newcommand{\INFNSEZIONEDIMILANOB}{INFN Sezione di Milano Bicocca and University, Milano, Italy}
\newcommand{\INFNSEZIONEDINAPOLIN}{INFN Sezione di Napoli, Napoli, Italy}
\newcommand{\INFNSEZIONEDIPADOVAA}{INFN Sezione di Padova and University, Padova, Italy}
\newcommand{\INFNSEZIONEDIPAVIAAN}{INFN Sezione di Pavia and University, Pavia, Italy}
\newcommand{\INFNSEZIONEDIPISAPIS}{INFN Sezione di Pisa, Pisa, Italy}
\newcommand{\IPSIINAF}{IPSI INAF Torino, Italy}
\newcommand{\UNIVERSITYOFPITTSBUR}{University of Pittsburgh, Pittsburgh, PA 15260, USA}
\newcommand{\UNIVERSITYOFROCHESTE}{University of Rochester, Rochester, NY 14627, USA}
\newcommand{\SLACNATIONALACCELERA}{SLAC National Accelerator Laboratory, Menlo Park, CA 94025, USA}
\newcommand{\SOUTHERNMETHODISTUNI}{Southern Methodist University, Dallas, TX 75275, USA}
\newcommand{\YORKUNIVERSITYTORONT}{York University, Toronto, Canada}

\affiliation{\ANL}
\affiliation{\BNL}
\affiliation{\CBPFCENTROBRASILEIRO}
\affiliation{\CERNEUROPEANORGANIZA}
\affiliation{\CIEMAT}
\affiliation{\CTI}
\affiliation{\VirginiaTech}
\affiliation{\IIS}
\affiliation{\CENTRODEINVESTIGACIO}
\affiliation{\CSU}
\affiliation{\Columbia}
\affiliation{\FNAL}
\affiliation{\INFNGSSILAQUILAITALY}
\affiliation{\INFNLNGSASSERGIITALY}
\affiliation{\INFNLNSCATANIAITALY}
\affiliation{\INFNSEZIONEDIBOLOGNA}
\affiliation{\INFNSEZIONEDICATANIA}
\affiliation{\INFNSEZIONEDIGENOVAA}
\affiliation{\INFNSEZIONEDIMILANOB}
\affiliation{\INFNSEZIONEDIMILANOM}
\affiliation{\INFNSEZIONEDINAPOLIN}
\affiliation{\INFNSEZIONEDIPADOVAA}
\affiliation{\INFNSEZIONEDIPAVIAAN}
\affiliation{\INFNSEZIONEDIPISAPIS}
\affiliation{\IPSIINAF}
\affiliation{\IIT}
\affiliation{\Imperial}
\affiliation{\INDIANINSTITUTEOFSCI}
\affiliation{\SaoJose}
\affiliation{\INSTITUTODINEURO}
\affiliation{\Lancaster}
\affiliation{\LANL}
\affiliation{\LSU}
\affiliation{\Holyoke}
\affiliation{\QueenMary}
\affiliation{\Rutgers}
\affiliation{\SLACNATIONALACCELERA}
\affiliation{\SOUTHERNMETHODISTUNI}
\affiliation{\Syracuse}
\affiliation{\TAMU}
\affiliation{\Tufts}
\affiliation{\Granada}
\affiliation{\Campinas}
\affiliation{\Alfenas}
\affiliation{\ABC}
\affiliation{\Bern}
\affiliation{\Palermo}
\affiliation{\UCL}
\affiliation{\UCSB}
\affiliation{\Chicago}
\affiliation{\Edinburgh}
\affiliation{\Florida}
\affiliation{\UNIVERSITYOFHOUSTONH}
\affiliation{\Kansas}
\affiliation{\Liverpool}
\affiliation{\Manchester}
\affiliation{\Michigan}
\affiliation{\Minnesota}
\affiliation{\NotreDame}
\affiliation{\Oxford}
\affiliation{\Penn}
\affiliation{\UNIVERSITYOFPITTSBUR}
\affiliation{\UNIVERSITYOFROCHESTE}
\affiliation{\Sheffield}
\affiliation{\Sussex}
\affiliation{\UTK}
\affiliation{\UTA}
\affiliation{\Warwick}
\affiliation{\YORKUNIVERSITYTORONT}

\author{P.~Abratenko}
\affiliation{\Tufts}
\author{N.~Abrego-Martinez}
\affiliation{\CENTRODEINVESTIGACIO}
\author{R.~Acciarri}
\affiliation{\FNAL}
\author{A.~Aduszkiewicz}
\affiliation{\UNIVERSITYOFHOUSTONH}
\author{F.~Akbar}
\affiliation{\UNIVERSITYOFROCHESTE}
\author{D.~Andrade~Aldana}
\affiliation{\IIT}
\author{L.~Aliaga-Soplin}
\affiliation{\UTA}
\author{F.~Abd~Alrahman}
\affiliation{\UNIVERSITYOFHOUSTONH}
\author{R.~Alvarez-Garrote}
\affiliation{\CIEMAT}
\author{C.~Andreopoulos}
\affiliation{\Liverpool}
% ...

%% file: abstract.tex
We present a deep neural net-based region of interest detection method (DNN ROI) for signal processing in the liquid argon time projection chambers of the Short-Baseline Neutrino (SBN) Program, SBND and ICARUS. DNN ROI addresses limitations of the traditional wire-by-wire thresholding algorithm by leveraging the full two-dimensional detector readout and cross-plane matching information. To account for detector performance variations, we explore training with augmented samples. We find that DNN ROI outperforms the traditional method in both low-level ROI identification performance and high-level reconstruction metrics for high-energy cosmic and accelerator neutrino interaction products, while also being more robust against detector variations, with or without sample augmentation.

%% file: Introduction.tex
The Short-Baseline Neutrino (SBN) Program~\cite{SBNProposal, SBNProgram} at Fermi National Accelerator Laboratory (FNAL) consists of multiple liquid argon time projection chamber (LArTPC) neutrino detectors~\cite{SBNProposal, ICARUSInagural, MicroBooNEInagural}. SBN has a broad program of experimental measurements and new physics searches utilizing neutrinos produced in the Booster Neutrino Beam (BNB)~\cite{BNB} and the Neutrinos at the Main Injector (NuMI) beam~\cite{NuMI} at Fermilab. 

LArTPCs reconstruct charged particle trajectories with high spatial resolution and precise calorimetry by imaging ionization electrons produced in charged particle tracks and showers. In a LArTPC, ionized electrons drift under a uniform electric field toward a series of wire planes at different orientations. The induction planes detect the induction current from ionization passing through the planes. The collection plane measures the current as the charge collects directly on the wires. The currents on the wires on each plane are digitized to form two-dimensional time-wire projection images. These images are used to reconstruct the three-dimensional trajectories of the charged particles produced in neutrino interactions. A critical early step in LArTPC event reconstruction is the signal processing chain, which extracts the reconstructed charge from these raw waveforms. This process is complicated by electronics noise, long-range induction effects, and the bipolar nature of induction-plane signals~\cite{WireCellSP}.

This work focuses on signal processing for the SBND
and ICARUS~\cite{ICARUSInagural} detectors, the near and far detectors of the SBN Program. The signal processing starts by removing noise that is coherent across adjacent groupings of wires read out by the same electronics board.
Following this, the waveforms are deconvolved to remove the effects of the field and electronics responses, resulting in approximately Gaussian-shaped charge pulses. Finally, the locations of these pulses are identified and recorded as regions of interest (ROIs), which isolate regions containing true charge signals, enabling efficient data processing and accurate downstream reconstruction. Only the portions of the deconvolved waveforms within ROIs are saved; these are then passed to a hit finding algorithm, which assigns position and shape information to the reconstructed charge pulses.

Traditional ROI detection uses thresholding and particle trajectory connectivity-based heuristics on deconvolved waveforms to identify the location of ionization signals~\cite{WireCellSP}. The traditional method is effective for most high-energy, sparse features such as isolated tracks. However, these heuristics face challenges for particular cases, such as extended trajectories and shower activity. They also can fail for complex charge depositions such as tracks that are oriented nearly perpendicular to the wire planes and produce highly prolonged charge signals with non-Gaussian shapes.

\begin{figure*}[t]
    \centering
    \includegraphics[width=\textwidth]{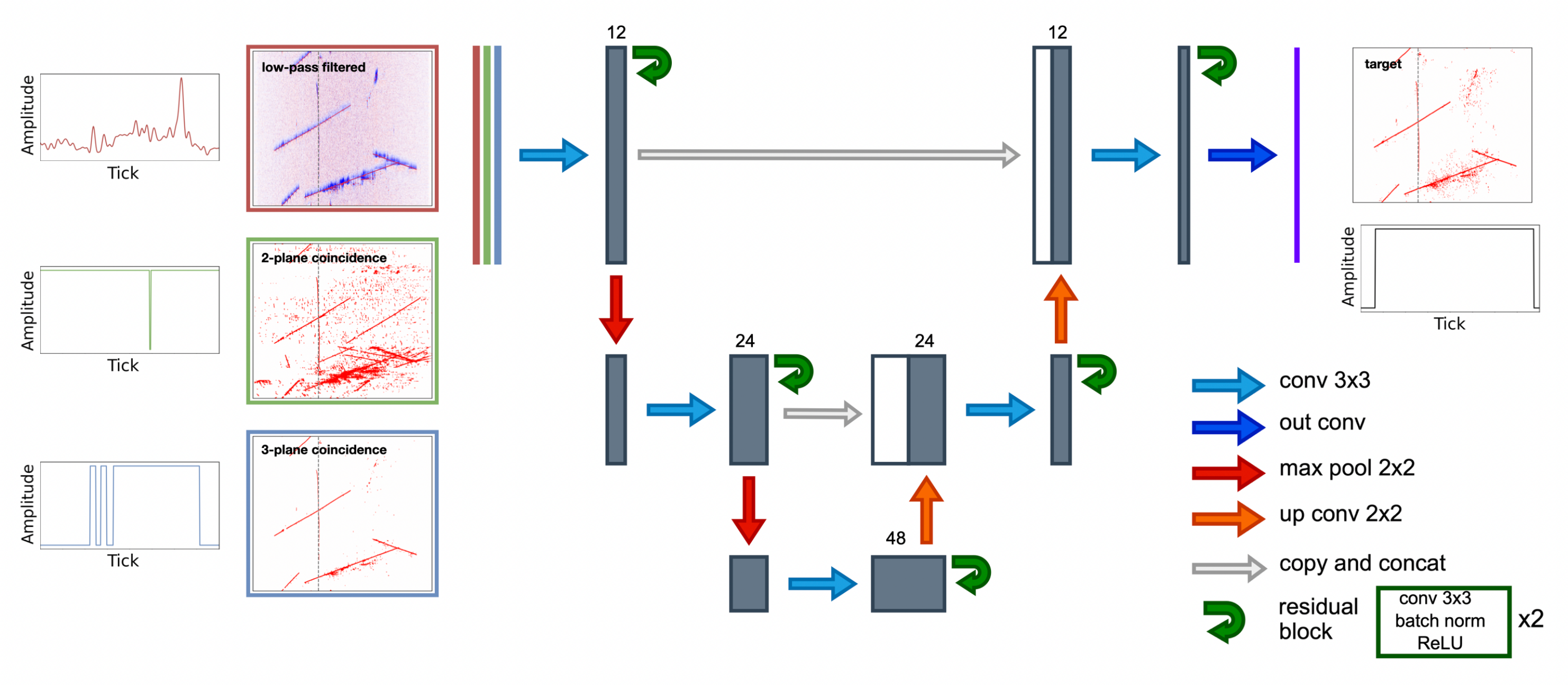}
    \caption{Structure of the DNN ROI network applied in SBND and ICARUS. The network applies a U-ResNet architecture~\cite{unet, uresnet1}, with three inputs (detailed in Sec.~\ref{sec:pre-process}) derived from the low-level signal processing. The objective of the network is to identify the location of true ionization depositions. Each layer consists of multiple $3\times3$ convolutions applied to each channel of the pixel map (the number of channels is denoted above each layer). In between layers, at the downsampling steps a max pool operation halves both pixel dimensions, while at the upsampling steps an up-convolution doubles both dimensions. Skip connections between the down-sampling and up-sampling stages are made through the copy and concat operation.}
    \label{fig:network}
\end{figure*}

To overcome these limitations, we employ a deep learning-based ROI detection method (DNN ROI), originally introduced in Ref.~\cite{DNNROI}. This method frames ROI detection as a 2D semantic segmentation task, labeling each pixel as either signal or noise. Crucially, it integrates geometric constraints across wire planes to enhance ROI fidelity, particularly for induction-plane signals affected by bipolar response cancellation. The structure of the network is displayed in Fig.~\ref{fig:network}. We apply the DNN ROI algorithm on the induction planes, where the challenge of signal identification is greatest.

In this work, we apply and extend our DNN ROI framework to the SBND and ICARUS detectors within the SBN Program. We explore various detector-specific optimizations, including filter tuning and augmented training strategies, to adapt DNN ROI to different noise environments and signal responses. Results from both low-level waveform comparisons and high-level reconstruction metrics demonstrate that DNN ROI improves signal identification efficiency and purity over traditional methods. These performance improvements are shown to be robust against different detector conditions.

A key focus of the application of DNN ROI in SBN is the verification that the networks are robust against realistic detector variations and defects observed in the data. Towards this goal, we demonstrate tests of the network performance against variations in the detector simulation informed by observations of the real detector performance. We also make augmentations of the training data such as removing random wires. This work can be understood as a case study in ``scientific robustness'' in machine learning. Namely, it addresses the need for neural networks to provide robust, unbiased results in the face of deformations to the data faced by a real experiment ~\cite{SciRobustNhan, SciRobust2, SciRobust3, AlexRobust}. The DNN ROI algorithm development for SBN both leverages ideas from this literature, and itself can serve as a testbed for ideas in the field. 

In addition, the application of DNN ROI in SBN is situated in the broader context of applications of machine learning (ML) in particle physics~\cite{NOvAML, CMSML, AntimatterML, MicroBooNEML, JetML, NoVAML2, SPINEML, MLReview, ArgoNeuTDNNROI, UboldiDNNROI}. In particular, the upcoming Deep Underground Neutrino Experiment (DUNE) will also use LArTPC neutrino detectors to study the properties of long-baseline neutrino oscillations, as well as search for proton decay and detect neutrinos from galactic supernovae~\cite{DUNETDR}. The algorithm development detailed in this work could be adapted for use in DUNE. The robustness studies, demonstrating resilience across varying detector conditions, offer insights that can inform the design and validation of future ML methods.

This paper is organized as follows. In Sec.~\ref{sec:pre-process}, we detail the pre-processing performed on the raw inputs to the network. In Sec.~\ref{sec:samples}, we describe the simulation sample generation and detector-specific pre-processing for ICARUS and SBND. Section~\ref{sec:network} details the network optimization, training, and inference pipeline. Section~\ref{sec:results} presents the evaluation metrics and results across various event topologies and noise conditions. We conclude in Sec.~\ref{sec:conclusion} with a summary and prospects for further integration of DNN ROI in future LArTPC workflows.

\technote{
\subsection{Technote Appendices}
For the technical note, we also include in appendices a variety of supplemental material relevant for documentation and details in the analysis. Appendix~\ref{app:software} details the software versions and samples applied in the results of this paper. Appendix~\ref{app:icarus-filter-optimization} discusses the procedure for optimizing the deonvolution filter values in ICARUS. (The optimization of the SBND deconvolution filter values will be described in a future note/publication.) The optimization of the network architecture is discussed in appendix~\ref{app:network-optimization}.
}

%% file: pre-process.tex
In both detectors, the raw ionization waveforms are processed to maximize the signal-to-noise ratio. The signal processing ultimately produces two outputs: an ROI waveform, which is used as an input to DNN ROI, and a charge extraction waveform. The ROIs are applied to the charge extraction waveform, which is taken as the output of the signal processing. 

In the first step of the signal processing, noise that is coherent across groupings of channels read out by the same electronics board is subtracted~\cite{ICARUSInagural, SBNDInagural}. Next, the signals are deconvolved consecutively in the time and wire directions~\cite{WireCellSP}.
In one dimension, a deconvolution forms an output signal $S(\omega)$ from an input $M(\omega)$ in the frequency domain as
\begin{equation}
    S(\omega) \equiv \frac{M(\omega)}{R(\omega)} F(\omega)\,,
    \label{eq:1Ddeconv}
\end{equation}
where $R(\omega)$ is the single electron response function and $F(\omega)$ is the filter function. In two dimensions, the deconvolution forms the output signal $S(\omega)$ as a function of a set of channels around each wire. This can be expressed in matrix form as~\cite{WireCellSP}
\begin{equation}
    S(\omega) \equiv \left(\vec{F}_w \cdot \bm{R}^{-1}(\omega) \cdot \vec{M}(\omega)\right)F(\omega)\,.
\end{equation}
In this equation, $\vec{M}(\omega)$ is the vector of waveforms on each channel around the central wire. $F(\omega)$ is the time-dimension filter function expressed in the frequency domain, as in Eq.~\ref{eq:1Ddeconv}. $\bm{R}$ is a symmetric matrix that specifies the signal response on each channel around the central wire, such that in matrix form it is equal to
\begin{equation}
    \bm{R}(\omega) = \begin{pmatrix}
        R_0(\omega) & R_1(\omega) & \ldots & R_{n-2}(\omega) & R_{n-1}(\omega)\\
        R_1(\omega) & R_0(\omega) & \ldots & R_{n-3}(\omega) & R_{n-2}(\omega)\\
        \vdots & \vdots & \ddots & \vdots & \vdots\\
        R_{n-2}(\omega) & R_{n-3}(\omega) & \ldots & R_{0}(\omega) & R_{1}(\omega)\\
        R_{n-1}(\omega) & R_{n-2}(\omega) & \ldots & R_{1}(\omega) & R_{0}(\omega)\\
    \end{pmatrix}\,,
\end{equation}
where $R_i(\omega)$ is the ionization signal response on a channel $i$ wires away. The matrix extends ten wires on either side of the central channel, for a total size of $21\times 21$. This matrix is constructed from the field response of charges approaching the wire plane, convolved with the response of the readout electronics~\cite{WireCellSP, ICARUSCalibration}.
Finally, $\vec{F}_w$ is the wire direction filter function. It does not depend on the frequency $\omega$.

We parameterize the wire filter function as a Gaussian, such that the $i$-th index of the vector is defined as
\begin{equation}
    \left(\vec{F}_w\right)_i \propto \exp\left[-\frac{1}{2}\left(\frac{i-i_0}{\sigma_w}\right)^2 \right]\,
    \label{eq:wirefilter}
\end{equation}
where $i_0$ is the index of the central wire and $\sigma_w$ is the width of the filter. We parametrize this filter in the frequency domain of the wire direction, i.e.~the wire frequency width of the filter, which is equal to $\pi/\sigma_w$. The function is normalized such that the amplitude of the measured signal ($\vec{M}(\omega)$) is preserved. 

We apply two time dimension filters for the two different objectives for the deconvolved waveform, ROI identification and charge extraction. The ROI identification waveform leverages a Wiener-like filter. In principle, given the signal response and noise power, one can define an optimal (``Wiener'') filter that maximizes the signal-to-noise ratio~\cite{Filtering}. However, in practice, the signal response in a LArTPC varies significantly depending on the ionization deposition pattern particle-to-particle. Thus, we define the filter function with a general functional form able to encapsulate a ``Wiener-like'' filter, and optimize the parameters in the filter functions using ionization depositions from simulated particle trajectories. We parameterize this filter as
\begin{equation}
\begin{split}
    F_W(\omega) \propto \exp\left[-\frac{1}{2}\left(\frac{\omega}{\sigma_W}\right)^a \right]\left(1 - \exp\left[-\left(\omega/f_c\right)^2\right]\right)\,,
    \label{eq:ROIfilter}
\end{split}
\end{equation}
where $\sigma_W$ is the width of the filter, $f_c$ is the low frequency cutoff, and $a$ is a constant that interpolates the Wiener-like filter from Gaussian-like ($a=2$) to step-function-like ($a\to\infty$). The Wiener-like filter maximizes the signal-to-noise ratio, at the expense of distorting the deconvolved waveform and making charge extraction challenging. Thus, a second filter is used to produce the waveform for charge measurement. This filter is given by
\begin{equation}
    F_Q(\omega) \propto \exp\left[-\frac{1}{2}\left(\frac{\omega}{\sigma_Q}\right)^2\right]\,,
    \label{eq:Qfilter}
\end{equation}
where $\sigma_Q$ is the width of the filter. As before, each of the filter functions are normalized such that the amplitude of the measured signal ($\vec{M}(\omega)$) is preserved. 

The values of the filter parameters in SBND and ICARUS are listed in Table~\ref{tab:filters}. Differences in the optimal filter parameters between the two detectors are largely due to differing noise conditions. SBND, unlike ICARUS, has its TPC readout electronics inside the cryostat, where the liquid argon cools the electronics, lowering the readout noise. Furthermore, in ICARUS there are longer cables in between signal amplification and digitization, which add a significant amount of noise.
As a result, in ICARUS it is optimal to cut more tightly in the frequency domain, only passing through the low frequencies where the signal power is dominant over noise. This difference results in broader deconvolved signals in the time domain in ICARUS than in SBND.

\begin{table}
    \centering
    \rowcolors{2}{gray!25}{white}
    \begin{tabular}{c | c | c}
    \rowcolor{gray!10}
    Filter Parameter & \makecell{SBND\\Value} & \makecell{ICARUS\\Value}\\
    \hline
    \multicolumn{3}{c}{$\vec{F}_w$ (Eq.~\ref{eq:wirefilter})}\\
    \hline
    Induction Wire Frequency ($\pi/\sigma_w$) & 1.05 $/\sqrt{\pi}$ & 0.4$/\sqrt{\pi}$ \\
    Collection Wire Frequency ($\pi/\sigma_w$) & 3.6 $/\sqrt{\pi}$ & 2.2$/\sqrt{\pi}$ \\
    \hline
    \multicolumn{3}{c}{$F_W(\omega)$ (Eq.~\ref{eq:ROIfilter})}\\
    \hline
    Front Ind.~Wiener-like Freq.~($\sigma_W$) [\si{\kilo\hertz}] & 150 & 65 \\
    Front Ind.~Wiener-like Power ($a$) & 5.5 & 4.4 \\
    Middle Ind.~Wiener-like Freq.~($\sigma_W$) [\si{\kilo\hertz}] & 150 & 65 \\
    Middle Ind.~Wiener-like Power ($a$) & 5.0 & 2.6 \\
    Collection Wiener-like Freq.~($\sigma_W$) [\si{\kilo\hertz}] & 250 & 70 \\
    Collection Wiener-like Power ($a$) & 3.0 & 3.4 \\
    Wiener-like Frequency Cutoff ($f_c$) [kHz] & 6 & 6\\
    \hline
    \multicolumn{3}{c}{$F_Q(\omega)$ (Eq.~\ref{eq:Qfilter})}\\
    \hline
    Gaussian Frequency ($\sigma_Q$) [\si{\kilo\hertz}] & 100 & 60 \\
    \end{tabular}
    \caption{Values of filter parameters in signal deconvolution, for both SBND and ICARUS. These parameters were individually optimized in SBND and ICARUS to maximize signal identification capabilities. Where specified, some parameters are different on each plane, or different between collection and induction planes. Otherwise, the parameter is the same on each plane.}
    \label{tab:filters}
\end{table}

\begin{figure}[t]
    \centering
    \includegraphics[width=0.49\textwidth]{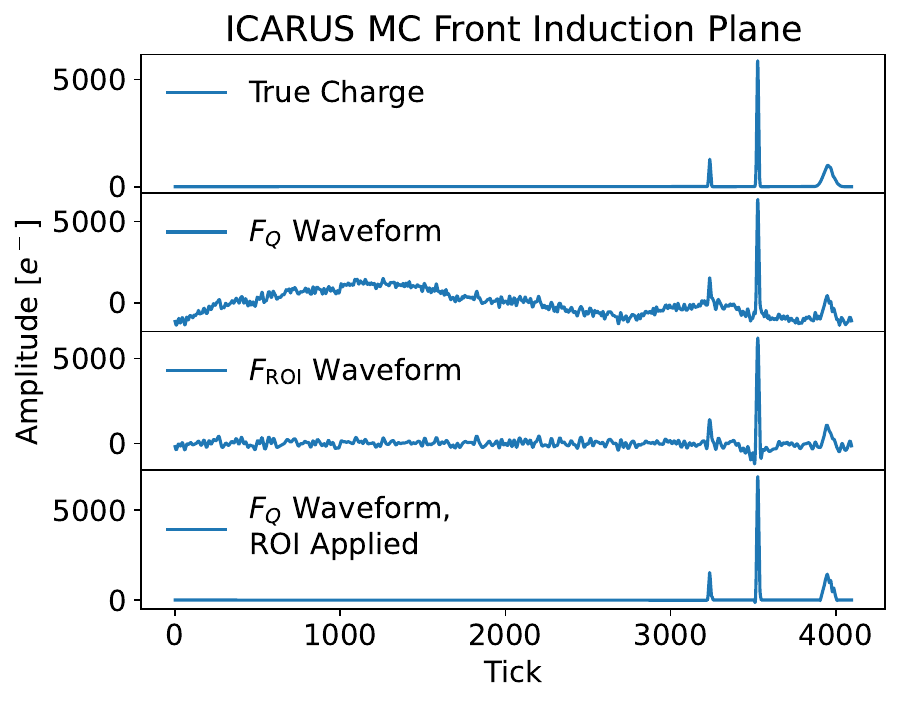}
    \caption{Example deconvolved waveform on the front induction plane in ICARUS Monte Carlo simulation, showing the impact of the Wiener-like (Eq.~\ref{eq:ROIfilter}) and charge extraction (Eq.~\ref{eq:Qfilter}) filters. The signals are induced by simulated cosmic muons. The charge extraction filter induces large noise fluctuations at low frequency, while the Wiener-like filter induces dips on either side of large signals, making both individually unsuitable for charge measurement. However, by applying the ROI onto the charge extraction waveform and subtracting a linear baseline, the charge can be suitably measured.}
    \label{fig:waveforms}
\end{figure}

The waveform from the Wiener-like filter is used to identify the windows where charge is present: ROIs. These ROIs are derived from a thresholding algorithm in the traditional case. For DNN ROI, the waveform is input to the neural network, which produces the ROIs as output. The ROI windows are applied as a mask on the charge extraction waveform, which is then recorded as the output of the signal processing. This process is depicted in Fig.~\ref{fig:waveforms}. The charge extraction waveform has significant low frequency oscillations. This is due to the fact that the signal power, especially on the induction planes, tends to zero near zero frequency; i.e., $\bm{R}^{-1}(\omega)$ tends towards infinity as $\omega\to0$. There is no low frequency cutoff in $F_Q$, and so any low frequency noise is made very large by the divergence in $\bm{R}^{-1}(\omega)$.
As shown in Fig.~\ref{fig:waveforms}, this feature is ameliorated by fitting a linear baseline on either side of the ROI, which is subtracted from the charge extraction waveform when the ROIs are applied. Thus, the objective of the ROI is to identify the presence of charge in a tight window so that low frequency oscillations do not impact its measurement. These features -- the charge identification efficiency and purity, as well as the charge measurement resolution and bias -- are all key metrics in the evaluation of DNN ROI performance.

\subsection{DNN ROI Input Channels}

As depicted in Fig.~\ref{fig:network}, DNN ROI network architecture takes three two-dimensional images (or frames) as input~\cite{DNNROI}. These frames are all derived from the output of the de-convolution. They are as follows:
\begin{enumerate}
    \item ROI Filter Output: This frame consists of the deconvolved signal on the wire plane, using the ROI filter (Eq.~\ref{eq:ROIfilter}). 
        
    \item Two-Plane (MP2) Coincidence: This frame encodes geometric constraints across wire planes. It identifies channels where activity in one plane coincides with activity in at least one other plane within a common time window on an overlapping wire. This information helps the network disambiguate true physics signals from noise and artifacts that are unlikely to be correlated across planes. The frame is a binary waveform that is equal to $1$ precisely when the coincidence condition is met, and $0$ otherwise. Signals over three times the noise RMS is considered for coincidence.
    
    \item Three-Plane (MP3) Coincidence: Similar to MP2, this frame highlights channels where signals are simultaneously present (in time) across all three wire planes, providing the strongest geometric constraint and a robust handle on true charge depositions. Signals over five times the noise RMS on the plane of interest, and signals over three times the noise RMS on the other planes are considered for coincidence.
\end{enumerate}

The target frame used during training is constructed by applying a charge threshold comparable to the noise RMS on the simulated true charge deposition.

%% file: SampleGen.tex
Training samples for DNN ROI are produced with SBND and ICARUS Monte Carlo simulation in the LArSoft software framework~\cite{LArSoft}. The primary particles in each event consists of neutrino interactions produced by the GENIE generator~\cite{GenieUserManual, GENIE} and cosmic rays simulated by CORSIKA~\cite{Corsika}.
Those particles are propagated through the detector with a GEANT4 simulation~\cite{GEANT4}. Finally, a detector simulation developed for the SBN program, inside the WireCell framework~\cite{WireCellSP}, produces the TPC response to ionization charge induced by energy depositions from particles propagating through the detector. This stage of the simulation is informed by the calibrations of both SBND and ICARUS to perform a realistic emulation of the detector response~\cite{SBNDInagural, ICARUSCalibration, ICARUSRecombination}.

To ensure the networks perform well across the full range of high-energy signal types encountered in SBN, we constructed a diverse training sample combining multiple physics-driven categories:
\begin{itemize}
    \item BNB $\nu$ + Cosmics: Booster Neutrino Beam (BNB) neutrino interactions with cosmic-ray muons. This sample represents the typical data observed in a readout of the SBND and ICARUS detectors.

    \item NuMI $\nu$ + Cosmics: Neutrinos at the Main Injector (NuMI) beam neutrino interactions with cosmic-ray muons. This sample represents the typical data in a readout of the NuMI beam in ICARUS. SBND is not in the path of the NuMI beam, and does not collect data from it.
    
    \item Prolonged Tracks with $\theta_{xz}$ ($\equiv \arctan\left(\vec{v}\cdot\hat{x}/\vec{v}\cdot\hat{z}\right)$, for a track direction $\vec{v}$) between 75$^\circ$--87$^\circ$: A dedicated sample of long, shallow-angle muon tracks, which are particularly challenging for induction-plane ROI finding due to bipolar cancellation and signal spread in time. Including these events in training is critical to improve the DNN’s ability to correctly reconstruct such track topologies. In data, such tracks typically come from cosmic muons, but can also be sourced by highly angled neutrino interactions.
    
    \item $\nu_e$ CC Interactions + Cosmics: A sample enriched in electromagnetic showers, such as those produced in $\nu_e$ charged-current (CC) interactions. This sample helps the network learn to preserve and enhance ROI finding in complex shower topologies, where traditional methods are sensitive to energy-dependent biases.
\end{itemize}
Given the DNN ROI’s primary focus on reconstructing GeV-scale neutrino interaction track and shower products, simulated radiogenic activity was not included in the training data.


We included a variety of data augmentations in these samples to reflect realistic effects in LArTPC detectors.
First, LArTPCs typically have a small number of non-responsive channels. To inform the network of this detector characteristic, bands of wires were randomly masked in the training input images. The position of the masked band was randomly selected on each wire plane, and the width of the number of wires to mask was sampled from a Gaussian, with a mean of 10 and a standard deviation of 5. This approach enables the network to learn the general feature of non-responsive wires, making it more versatile across different regions of the detector. This augmentation was applied to both ICARUS and SBND samples.

\begin{figure*}[]
    \centering
    \includegraphics[width=\linewidth]{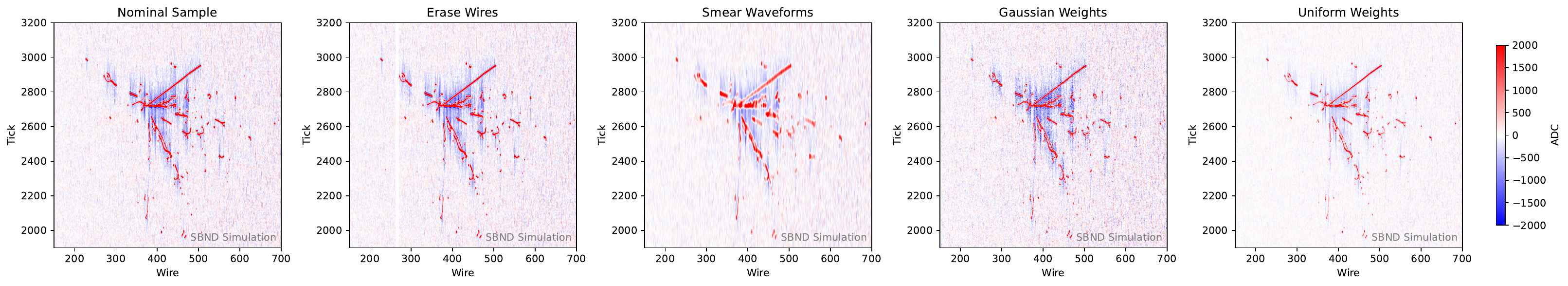}
    \caption{Demonstration of image augmentations used to simulate detector variations on the middle induction plane in SBND. From left to right: nominal image, image with a randomly masked band of wires, image smeared in the tick direction using a Gaussian kernel, image with pixel-wise random scaling, and image with event-wise random scaling applied to all pixel values. The magnitude of the variation is exaggerated compared to what is used for training for illustrative purposes.}
    \label{fig:sbnd_sample_var}
\end{figure*}

In addition, the samples also contained a number of augmentations specific to the two detectors, determined by their observed performance in data taking. These detector-specific augmentations are detailed in the following sections~\ref{subsec:ICARUS-sample} and~\ref{subsec:SBND-sample}.
The distributions of the detector variations are summarized in Tab.~\ref{tab:omnidetector}, and example images are shown in Fig.~ \ref{fig:sbnd_sample_var} and Fig.~\ref{fig:icarus_sample_var}. 

\subsection{ICARUS Sample Augmentation}
\label{subsec:ICARUS-sample}

A key focus of the ICARUS sample generation was to encapsulate the variations in the ICARUS detector observed over its operation. There are many effects in the ICARUS TPC that must be taken into account for a realistic assessment of the performance of the detector, such as the drift electron lifetime, transparency of the induction planes, and level of the coherent and incoherent noise on each channel~\cite{ICARUSCalibration}. These effects all vary, either over the runtime of the experiment or spatially across the detector, or both. By integrating the variation of these effects into the ICARUS DNN ROI samples, we ensured that the network would not overtrain on a particular simulation of the detector performance. 

In ICARUS, we produced two versions of training samples. In the ``Nominal'' sample,
detector simulation parameters were fixed at their nominal values. In the ``OmniDetector'' sample, every 10 events were simulated with a different set of randomly thrown simulation parameter values. The varied parameters impact the noise level, channel gain, and ionization signal shape on the TPC wire planes. These variations are detailed in Tab.~\ref{tab:omnidetector}. Ten parameters are varied in total. We determined the range of these parameters from considerations on the observed performance of the ICARUS detector~\cite{ICARUSCalibration}.

\begin{table*}
  \caption{Parameter distributions used for detector variations. \textit{Unif} denotes a uniform distribution, and $N$ denotes a normal distribution.}
  \label{sample-table}
  \centering
  \begin{tabular}{lll@{\hspace{1.5em}}l}
    \toprule
    Detector   &  Parameter     & Nominal Value  & Variation Distribution \bigstrut\\
    \midrule
    \multirow{6}[2]{*}{ICARUS} & Coherent Noise Scale & 1 & $N(1,0.05^2)$ \bigstrut[t]\\
    & Incoherent Noise Scale & 1 & $N(1,0.05^2)$\\
    & Electron Lifetime & \SI{5}{\milli\second} & \textit{Unif}(2,10) \SI{}{\milli\second}\\
    & Relative Gain (per-plane) & 1 & $N(1,0.1^2)$\\
    & Shaping Time (per-plane) & \SI{1.3}{\micro\second} & $N(1.3,0.05^2$) \SI{}{\micro\second}\\
    & Middle Ind.~Signal Shape & Bin 7 & Bins \textit{Unif}(1-15) \bigstrut[b]\\
    \midrule
    \multirow{2}[2]{*}{SBND} & Smear Waveforms & 0 & $N(1,2^2) \text{ ticks} $ \bigstrut[t]\\
    & Pixel weight & 1 & $N(1,0.05^2)$ \bigstrut[b]\\
    \midrule
    Common & Masked Wire Band Width & 0 & $N(10,5^2)$ wires \\
    \bottomrule
  \end{tabular}
  \label{tab:omnidetector}
\end{table*}

\begin{figure}[t]
    \centering
    \includegraphics[width=0.49\textwidth]{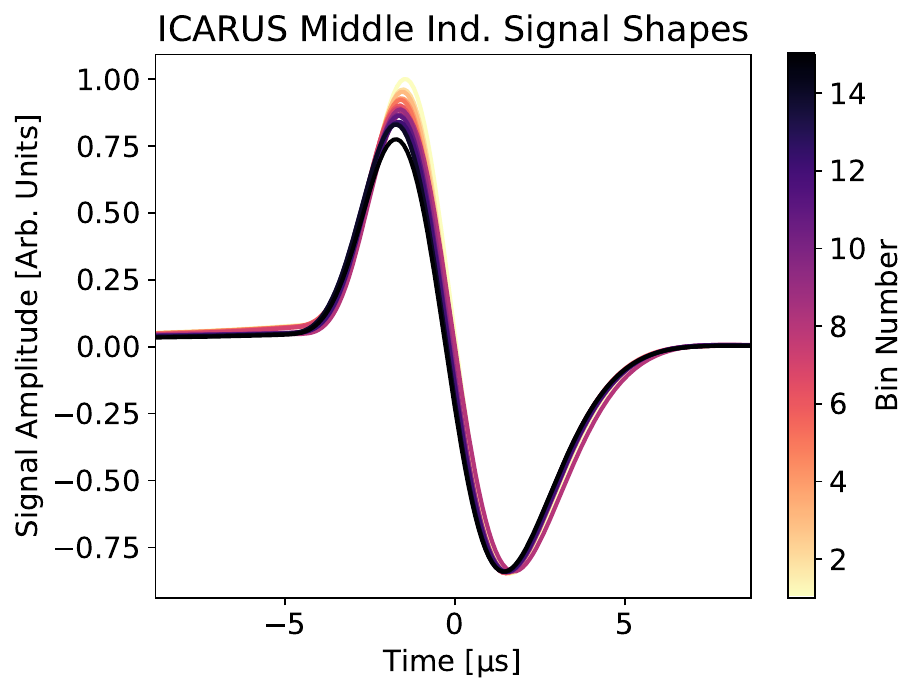}
    \caption{Variation in the single electron response across the middle induction wire plane in the ICARUS TPCs. The response is shown in 15 bins of charge intransparency in the detector. The signal amplitude is normalized to 1 for the most intransparent (bin 1) signal.}
    \label{fig:middle-ind-signal-shapes}
\end{figure}

One feature of ICARUS that is not introduced in Ref.~\cite{ICARUSCalibration} but included here is the variation in the middle induction signal shape across the detector. This feature is due to a variable intransparency to charge across the middle induction wire planes in each TPC. This charge intransparency has the effect of inducing a collection-like unipolar pulse in the middle induction plane signal response~\cite{Intransparency1, Intransparency2, Intransparency3}. It increases the magnitude of the up-peak in the middle induction signal shape, while interfering destructively with the down-peak. The effect of the variable intransparency on the middle induction signal shape across the ICARUS TPC is shown in Fig.~\ref{fig:middle-ind-signal-shapes}.

\begin{figure*}[]
    \centering
    \includegraphics[width=\linewidth]{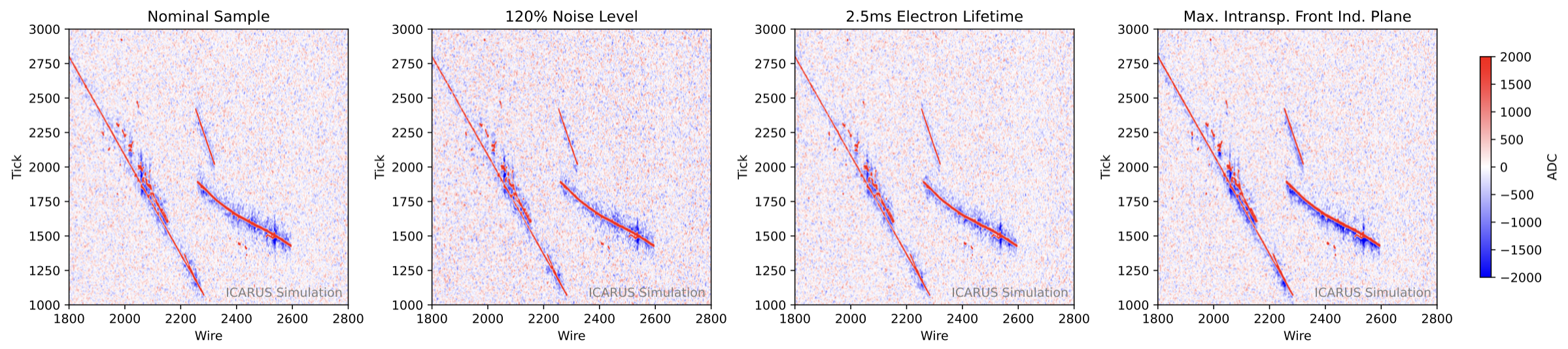}
    \caption{Demonstration of image augmentations used to simulate detector variations on the middle induction plane in ICARUS. From left to right: nominal image, image with 120\% noise level, image with \SI{2.5}{ms} electron lifetime, and image with maximal intransparency on the front induction plane.}
    \label{fig:icarus_sample_var}
\end{figure*}

The impact of each detector variation on an example frame in ICARUS is shown in Fig.~\ref{fig:icarus_sample_var}.



\subsection{SBND Sample Augmentation}
\label{subsec:SBND-sample}

SBND experiences similar types of detector effects as ICARUS, but to a much lesser extent. The detector variations observed in SBND are not large enough to impact cross-plane coincidence meaningfully, but they can alter the waveforms in the ROI filter output images from those expected from simulation. Taking this into account, detector variation samples for SBND were produced by directly augmenting the ROI filter output image to emulate the impact of detector variations, rather than fully regenerating the simulation-based variation samples as done for ICARUS.  

Most detector effects in LArTPCs manifest in ways analogous to image distortions. For example, electron diffusion can appear as smeared images, while noise or gain variations can shift the pixel values based on the position-dependent features. To simulate smearing effects, input images were convolved with a Gaussian kernel in the time tick direction, where the standard deviation was drawn from a normal distribution corresponding to $N(1,2^2)$ ticks\footnote{$N(\mu, \sigma^2)$ denotes a normal distribution with mean $\mu$ and standard deviation $\sigma$}. To simulate scaling effects, image pixel values were modified by two approaches. First, pixel-wise scaling, in which each pixel was assigned an independent weight. And second, event-wise scaling, where a single weight was applied uniformly across the whole image. In both cases, the weights were sampled from a Gaussian distribution, $N(1,0.05^2)$. The impact of each detector augmentation on an example frame in SBND is shown in Fig.~\ref{fig:sbnd_sample_var}.

%% file: Network.tex
\label{sec:network}
For application to SBN detectors, computational efficiency is a critical requirement. The trained networks need to be able to run on CPUs with reasonable memory and time requirements to be applied in data processing.


To satisfy these conditions, we adopted a strategy of reducing the input image size through ``chunking'' and downsampling. Chunking here means splitting an image into a set of smaller arrays (``chunks'') along a dimension.
Along the tick axis, we apply downsampling by averaging over fixed intervals of ticks, effectively reducing the resolution in the tick dimension. To maintain consistency with the digitized resolution of the cross-plane inputs, the waveform was downsampled by a factor of 4, the time window size used for identifying signal coincidence across the planes. Along the wire axis, images are split into smaller chunks to allow for inference to run on smaller images, which are then directly placed next to each other to reconstruct the full image. No performance degradation was observed near the chunk boundaries -- ROI identification performance and higher-level reconstruction outputs showed no artifacts or loss in the edge regions where the stitching occurs. In general, such resizing preserves the primary task, which is to learn the general structure of signal ROIs, rather than specific local features. However, this does change the effective resolution within the image, affecting the significance of small-sized energy depositions or noise. Therefore, maintaining a consistent configuration for downsampling and chunking between training and inference is important for optimal performance. In ICARUS, a downsampling factor of 4 is applied, with 2 chunks for the first induction plane (1024$\times$1056 pixels) and 4 chunks for the second (1024$\times$1400 pixels). In SBND, a downsampling factor of 4 with 2 chunks is used, resulting in 857$\times$992 pixels for both induction planes. Each wire plane in both detectors has its own network.

\begin{figure}[]
    \centering
    \includegraphics[width=0.49\textwidth]{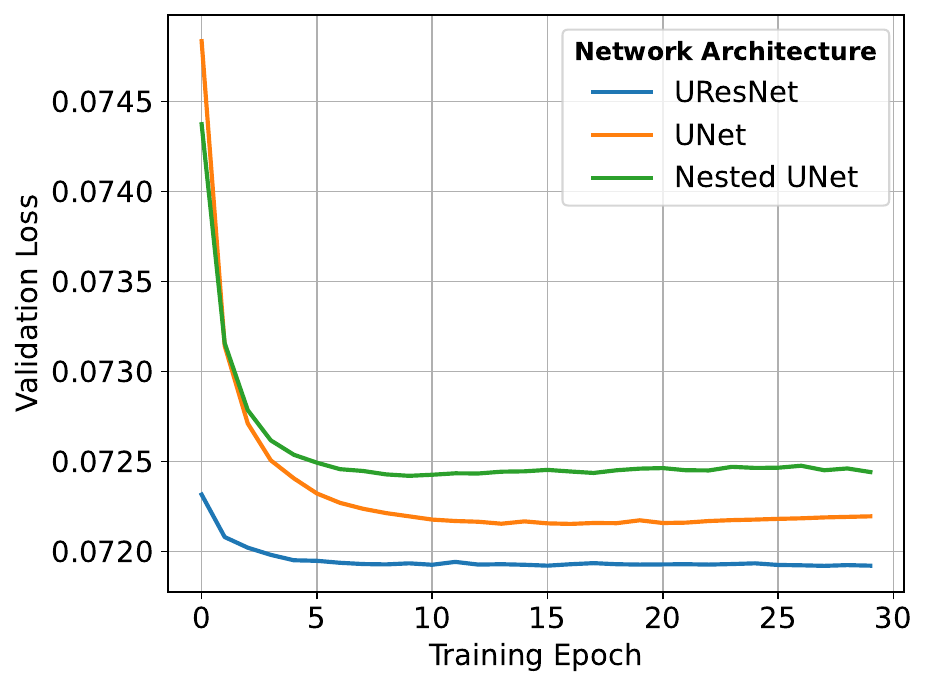}
    \caption{Validation loss for three different optimized model architectures trained on ICARUS images.}
    \label{fig:learning_curve}
\end{figure}

ROI finding can be approached as a binary image segmentation task. We explored variants of the U-Net neural network architectures, widely used for segmentation tasks, using the 3 input images as 3 input channels. Three network architectures were tested: U-Net~\cite{unet}, U-ResNet~\cite{uresnet1, uresnet2, uresnet3}, and Nested U-Net~\cite{nestedunet}. For network optimization, we performed a hyperparameter scan across multiple variables, including learning rate, optimizer type, and optimizer-specific parameters. Each network architecture was optimized, then compared using the Dice-S{\o}rensen coefficient of validation samples~\cite{Dice1945, Sorensen1948, Zijdenbos1994}, which quantifies the similarity between the predicted and target images, defined as
\begin{equation}
    \mathrm{DSC}(P,G) \;=\; 
\frac{2 \sum_{i=1}^{N} p_i g_i}
     {\sum_{i=1}^{N} p_i \;+\; \sum_{i=1}^{N} g_i},
    \label{eq:DSC}
\end{equation}
where $p_i$ is the predicted ROI probability for pixel $i$, $g_i$ is the binary target truth for pixel $i$, and $N$ is the total number of pixels in the image. A fixed threshold of 0.5 was applied to the network output logits to classify each pixel as either ROI or non-ROI. For both SBND and ICARUS, the best performing architecture was U-ResNet, a U-Net architecture with downsampling encoder blocks and upsampling decoder blocks, with the residual blocks from ResNet~\cite{resnet} in place of the convolutional blocks. The SBND model was trained with the stochastic gradient descent optimizer with learning rate of 0.01, momentum of 0.9, and weight decay of 0.0005. The ICARUS model was trained with the ADAM optimizer~\cite{adam} with a learning rate of 0.001, and the algorithm hyperparameters ($\beta_1$, $\beta_2$) = (0.9, 0.999). Figure \ref{fig:learning_curve} shows the validation loss curve for different network architectures, each with optimized hyperparameters, tested on ICARUS samples. 


Once a model with optimal performance was chosen, pruning was applied by gradually reducing the number of hidden units in each layer until the performance began to degrade. This optimization speeds up the inference time, by reducing the total number of operations.

Due to the sparse signal distribution of LArTPC data, input images have a significant class imbalance between ROI and non-ROI pixels. To address this, the binary cross-entropy loss was constructed as:

\begin{equation}
\mathcal{L} = - \frac{1}{N} \sum_{i=1}^{N} \Big[ w_s \, y_i \log \hat{y}_i + w_b \, (1 - y_i) \log (1 - \hat{y}_i) \Big],
\end{equation}
where \(y_i\) is the true label, \(\hat{y}_i\) is the predicted probability, \(w_s = 9\) is the ROI weight, \(w_b = 1\) is the non-ROI weight, and \(N\) is the total number of pixels. This weighting helps ensure that signal regions are preserved during training, and was optimized by scanning values and evaluation of metrics disussed in Sec. \ref{sec:results}.


Models were implemented in PyTorch~\cite{paszke2019pytorch}, and training was conducted on NVIDIA A100 GPUs with \SI{40}{GB} memory. The inference was executed within the LArSoft framework~\cite{LArSoft}, via the C++ based \textit{libtorch} library. Inference is run on CPUs due to computing availability, and takes \SI{30}{s} wall time and \SI{3}{GB} memory for SBND and \SI{700}{s} and \SI{8}{GB} for ICARUS per an event across all wire planes. 


%% file: Results.tex
\label{sec:results}
For both SBND and ICARUS, DNN ROI led to significant improvements over the traditional thresholding algorithm in ROI finding and ionization charge extraction fidelity. The performance of the network was quantified for pixels and ROIs using the $\mathrm{Efficiency}\times\mathrm{Purity}$ as a figure-of-merit, where these are given as
\begin{equation}
\begin{split}
\label{eq:roi_pixel_metrics_dsc_style}
\mathrm{Pixel\;Efficiency} &=
\frac{\sum_{i=1}^{N} p_i g_i}{\sum_{i=1}^{N} g_i}\qquad \\
\mathrm{Pixel\;Purity} &=
\frac{\sum_{i=1}^{N} p_i g_i}{\sum_{i=1}^{N} p_i}\\[1mm]
\mathrm{ROI\;Efficiency} &=
\frac{1}{R} \sum_{r=1}^{R} \min\Big(1,\sum_{i \in \mathcal{R}_r} p_i \Big)\qquad \\
\mathrm{ROI\;Purity} &=
\frac{1}{\hat{R}} \sum_{r=1}^{\hat{R}} \min \Big(1,\sum_{i \in \hat{\mathcal{R}}_r} g_i \Big),
\end{split}
\end{equation}
where $p_i$ and $g_i$ are defined as in Eq.~\ref{eq:DSC}.

Pixel-level efficiency and purity are computed by summing over all pixels, while ROI-level metrics consider contiguous regions of pixels. Each contiguous region of true pixels defines a true ROI $\mathcal{R}_r$, and each contiguous region of predicted pixels with probability greater than 0.5 defines a predicted ROI $\hat{\mathcal{R}}_r$. $R$ and $\hat{R}$ are the total numbers of true and predicted ROIs, respectively. Although the product of purity and efficiency is reported as the figure-of-merit, each were separately improved for all cases shown.

In addition, we investigated the impact of the improved ROI finding on higher-level information. For instance, shower completeness is defined as 

\begin{equation}
\label{eq:FOM}
\mathrm{Shower\;Completeness} \;=\;
\frac{\displaystyle \sum_{i \in \mathrm{ROI}} Q_i^{\mathrm{true}}}
     {\displaystyle \sum_{i \in \mathrm{Shower}} Q_i^{\mathrm{true}}} \,,
\end{equation}
where $Q_i^{\mathrm{true}}$ is the true deposited charge in pixel $i$.

Signal processing performance strongly depends on the shape of the ionization signals, which is determined by the relative orientation of the ionizing particle which produced the signal with respect to the wire planes. We therefore parameterize many of the results of this section in terms of the angles of the ionizing particle trajectories. We use a right-handed detector coordinate system: $x$ is the direction of the electron drift, $y$ is vertical upward direction, and $z$ is the direction of the neutrino beam. Further details on the two detectors are available in Ref.~\cite{ICARUSInagural, SBNDInagural}.

For all of the results shown here, we leverage samples from validation datasets that are independent of the original training dataset. In the section, we detail which type of sample is utilized alongside each relevant result.

\subsection{ICARUS Results}
DNN ROI enhances the ICARUS ROI idenfitication performance, especially for prolonged tracks and electromagnetic showers. 
The network demonstrates greater robustness against detector simulation variations in its performance than the traditional ROI algorithm. Notably, this improved robustness is demonstrated even when the network is not trained on data including such variations in the detector simulation.

\begin{figure}[]
    \centering
    \includegraphics[width=0.49\textwidth]{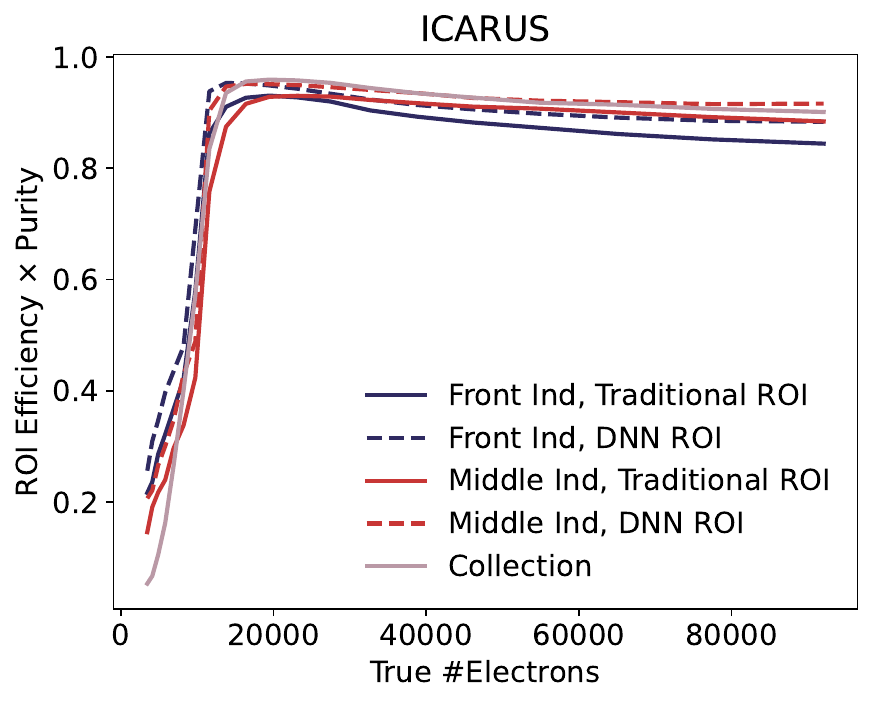}
    \includegraphics[width=0.49\textwidth]{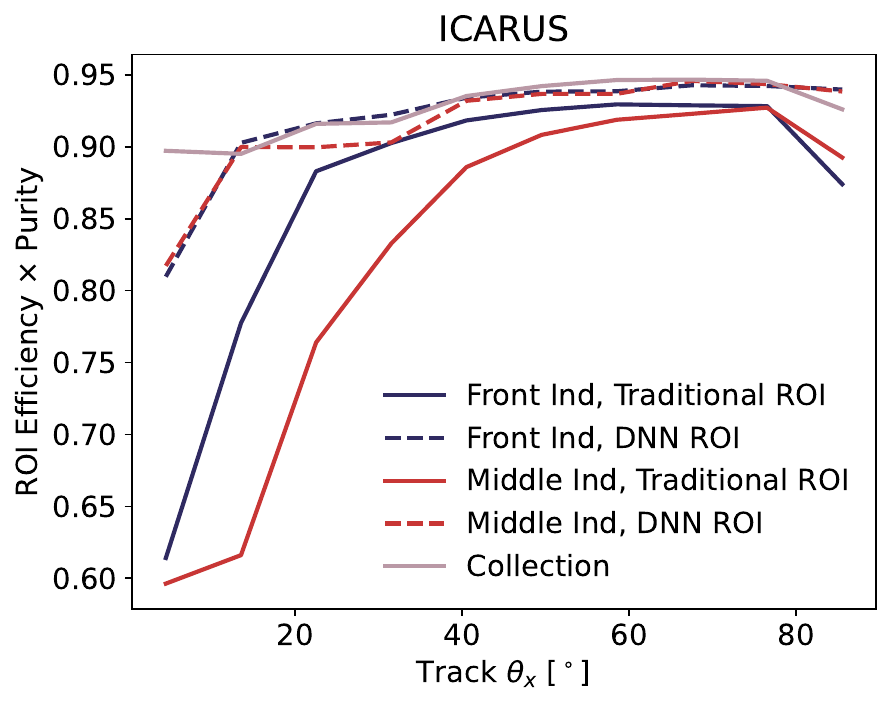}
    \caption{Comparison of performance of traditional and DNN ROI performance for track-like charge identification in ICARUS. This result utilizes a sample of simulated BNB neutrinos with cosmic-ray activity included. Shown as a function of the true number of electrons in the track hit (top), as well as by the track angle to the drift electric field ($\hat{x}$) (bottom). In the track angle plot, a threshold of $10^4$ electrons is applied when calculating the efficiency (see Eq.~\ref{eq:roi_pixel_metrics_dsc_style}).}
    \label{fig:icarus-track-perf}
\end{figure}
Figure~\ref{fig:icarus-track-perf} shows the performance of ROI identification for track-like charge depositions (such as muons and protons) on each wire plane. This result utilizes a sample of simulated BNB neutrinos with cosmic-ray activity included. The performance is shown with respect to the track angle $\theta_x \equiv \arccos\left(\hat{x}\right)$, where $\hat{x}$ is the component of the track direction along the drift electric field. A threshold on the number of electrons in the true deposition is applied where specified. On the induction planes, the performance of the traditional and the DNN ROI method is compared. On the collection plane, only the traditional algorithm is used. The set of tracks is taken from a sample of BNB neutrino events with simulated cosmic rays. The efficiency of ROI identification is improved on both induction planes, especially at low track angle to the drift electric field where the ionization signal is highly extended along the drift direction. The increased efficiency is attained while also maintaining a higher ROI purity, resulting in the significant improvement in the ROI Efficiency$\times$Purity metric shown in the figure.

\begin{figure}[t]
    \centering
    \includegraphics[width=0.49\textwidth]{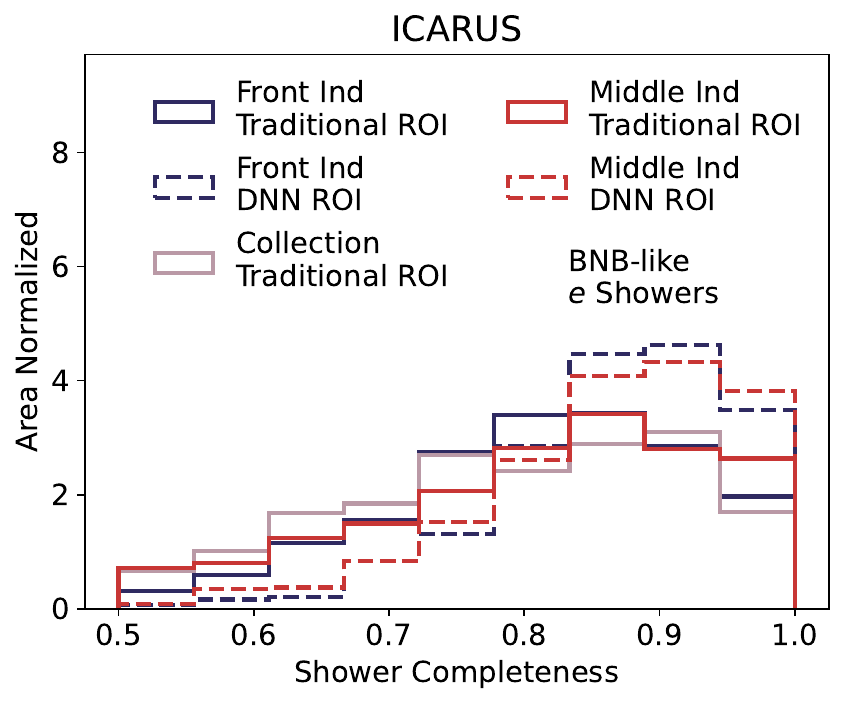}
    \caption{Comparison of the performance of traditional and DNN ROI performance for shower charge identification in ICARUS. Each entry in the histogram consists of one simulated shower from a distribution of electron showers similar to those expected from $\nu_e$ CC interactions in the BNB.}
    \label{fig:icarus-shower-perf}
\end{figure}

Figure~\ref{fig:icarus-shower-perf} shows the performance of ROI identification on each wire plane for shower-like charge depositions from electrons simulated with the spectrum produced by BNB $\nu_e$ charged-current interactions. DNN ROI identification improves the shower completeness on both induction planes. Although not shown in the figure, the purity of ROIs is also separately improved. 

To evaluate the robustness of DNN ROI identification against detector variations, we generated dedicated validation samples by simulating key variations observed in the detector. These include high noise, low electron lifetime, and increased wire intransparency. These samples are included in addition to the Nominal and OmniDetector samples introduced in Sec.~\ref{subsec:ICARUS-sample}. Each additional sample represents a particularly challenging state of the detector with respect to the nominal configuration. The added variations are extreme with respect to the operation of the ICARUS detector, and should therefore be considered on the edge of the known variations. For example, the low electron lifetime applies a value of \SI{2.5}{\milli\second}. This value attenuates the charge signal by up to 33\%, and is lower than any value included in the ICARUS physics data~\cite{ICARUSCalibration}.

\begin{figure}[]
    \centering
    \includegraphics[width=0.49\textwidth]{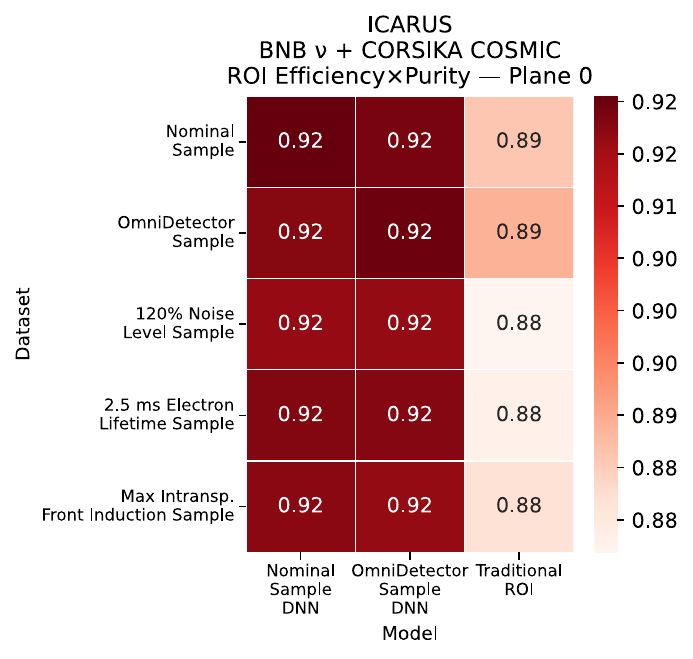}
    \caption{ROI Efficiency$\times$Purity for Plane 0 (first induction plane) in ICARUS under different detector variations, comparing traditional ROI and DNN ROI models. Shown for a sample of simulated BNB neutrinos with cosmic-ray activity included.}
    \label{fig:icarus-var-bnb-P0}
\end{figure}

\begin{figure}[]
    \centering
    \includegraphics[width=0.49\textwidth]{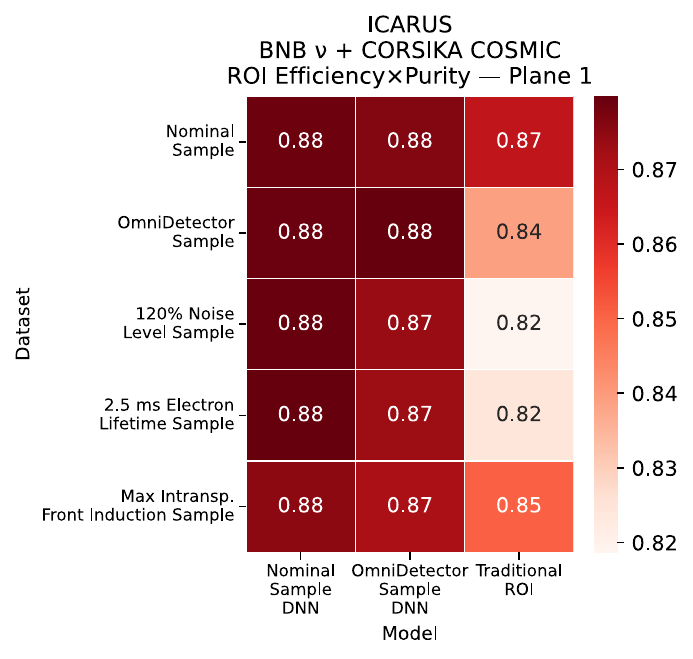}
    \caption{ROI Efficiency$\times$Purity for Plane 1 (middle induction plane) in ICARUS under different detector variations, comparing traditional ROI and DNN ROI models. Shown for a sample of simulated BNB neutrinos with cosmic-ray activity included.}
    \label{fig:icarus-var-bnb-P1}
\end{figure}

Figures~\ref{fig:icarus-var-bnb-P0} and ~\ref{fig:icarus-var-bnb-P1} show the ROI Efficiency$\times$Purity metric for the front induction and middle induction planes respectively, across these different detector variations, comparing traditional ROI finding to DNN ROI models trained on both Nominal and OmniDetector training samples. 

On the front induction plane, these variations only have a small impact on the performance of the traditional ROI algorithm. From this result, it can be inferred that the objective of ROI identification on the front induction plane is largely not sensitive to even extreme variations in the performance of the ICARUS TPC. Both networks display the same level of stability in their performance as the traditional algorithm.

On the middle induction plane, the same variations meaningfully impact the performance of the traditional ROI algorithm. The most challenging sample for the traditional ROI algorithm is the variation where the noise level is increased by 20\%. This sample obtains a 7\% lower value in the ROI Efficiency$\times$Purity metric relative to the nominal detector sample. Despite this, the performance of both DNN ROI models are unaffected by any of the detector model variations. This result is especially notable for the 20\% high noise sample, which is far outside the Nominal detector simulation and is also not well-represented in the OmniDetector sample. It is four standard deviations outside the average simulated noise scale (see Table~\ref{tab:omnidetector}).

We hypothesize that this result may be due to the fact that the network can learn charge topology and correlations across planes through cross-plane matched inputs. These features may be inherently more robust against detector effects than the signal-to-noise thresholding applied by the traditional algorithm~\cite{WireCellSP}. Furthermore, there is a significant variation in the topology and magnitude of charge depositions from different particle types, energies, and orientations. These variations, present in each training sample, may already represent a broader set of signals than are induced by realistic (or even extreme) variations in the detector simulation.

\technote{
\subsubsection{Technote Only: Future Work}
The development of DNN ROI in ICARUS has also highlighted a few challenges in the signal processing that will need to be addressed in future work on the experiment. 

First, the recovery of high-angle tracks itself presents a challenge to downstream high-level reconstruction. Such tracks typically have extended waveforms with a significant number of stochastic fluctuations, and are not well fitted by a single Gaussian shape. The hit finding reconstruction typically assumes charge depositions to have such a Gaussian shape. For these extended waveforms, it fits a ``pulse train'' of Gaussian hits (up to 25) that together can model the extended pulse. However, such a large number of hits creates significant combinatorics for the Pandora pattern matching reconstruction. Occasional events will have particular tracks that can take hours to days for Pandora to resolve. To remedy this, we have disabled hit trains for input to Pandora. Such extended pulses are instead (poorly) fit to a single Gaussian. However, this may not be an optimal solution, and this issue merits further investigation.

Second, the increase on completeness for shower-like depositions is accompanied by an increase in fake hits from low frequency oscillations (this is especially the case for showers that are extended on individual waveforms). These low-frequency-oscillation ``hits'' pollute the shower charge reconstruction on the induction planes and mitigate the improvement in shower charge resolution that would otherwise be obtained from the increase in completeness. Further work is needed to discriminate shower charge from low frequency hits, either through further signal processing development, or development of the high-level reconstruction.
}

\subsection{SBND Results}

\begin{figure}[t]
    \centering
    \includegraphics[width=0.5\textwidth]{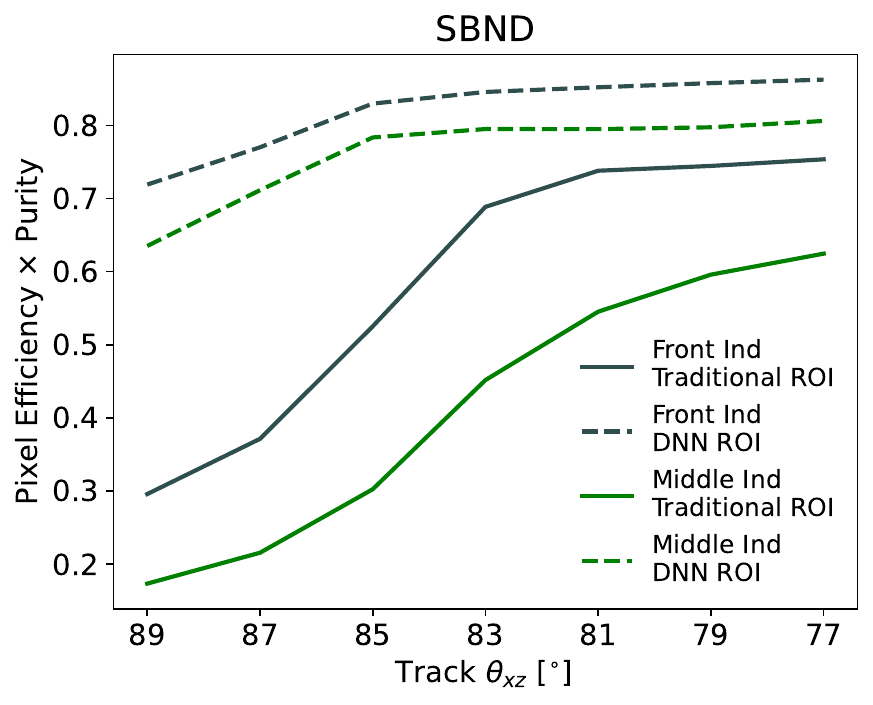}
    \caption{Pixelwise ROI identification efficiency$\times$purity metric evaluted on prolonged muon tracks.}
    \label{fig:sbnd-pixel_eval}
\end{figure}

Fig.~\ref{fig:sbnd-pixel_eval} shows the ROI-finding performance on prolonged tracks in the induction planes in SBND, comparing DNN ROI to the traditional method. The pixel efficiency and purity are calculated without applying a threshold on the deposited charge. The performance is shown with respect to the track angle $\theta_{xz}$. DNN ROI shows significant improvement across all angle ranges, with less loss in performance for the most prolonged angle bins compared to the traditional ROI finding. For both methods, the middle induction plane shows reduced performance relative to the front induction plane due to its more symmetric bipolar signal response. However, DNN ROI decreases the performance gap between the two induction planes.
Note that the performance of SBND in Fig.~\ref{fig:sbnd-pixel_eval} should not be directly compared to that of ICARUS in Fig.~\ref{fig:icarus-track-perf}. Identifying each wire-time pixel of charge is a more challenging requirement than identifying some fraction of a deposition within an ROI.

\begin{figure}[t]
    \centering
    \includegraphics[width=0.49\textwidth]{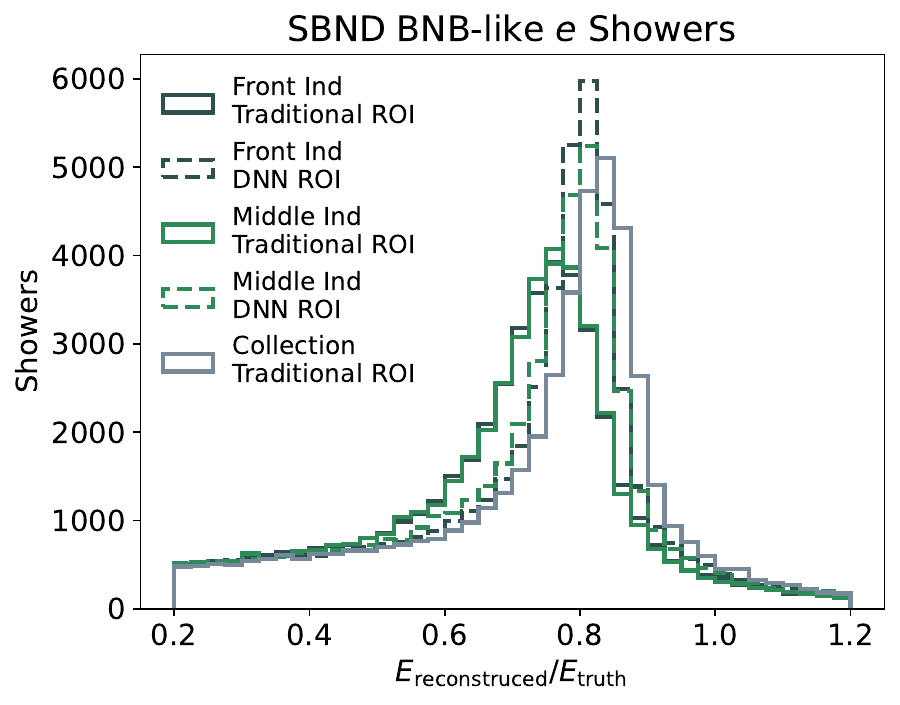}
    \caption{Ratio of the reconstructed electron energy to the simulation truth electron energy, for BNB-like electron neutrino events.}
    \label{fig:sbnd-shower-perf}
\end{figure}

Figure~\ref{fig:sbnd-shower-perf} shows the energy reconstruction performance for neutrino-induced electron showers. On both induction planes, energy reconstruction improves substantially with DNN ROI, confirming the positive impact of improved ROI finding and charge extraction performance on physics reconstruction. The observed improvement in shower energy reconstruction can be attributed to two main factors. First, shower-like energy depositions, especially higher-energy ones, often consist of multiple smaller branching trajectories, as can be seen in the example of Fig. \ref{fig:sbnd_sample_var}. Branches oriented at prolonged angles experience signal attenuation due to the bipolar response. DNN ROI is more effective at recovering such signals than the traditional method, similar to the case of prolonged muons. Secondly, showers frequently produce multiple small and locally isolated charge depositions. More effective identification of these lead to improved reconstruction of the total shower energy. 

\subsection{Discussion}

\begin{figure}[t]
    \centering
    \includegraphics[width=0.45\textwidth]{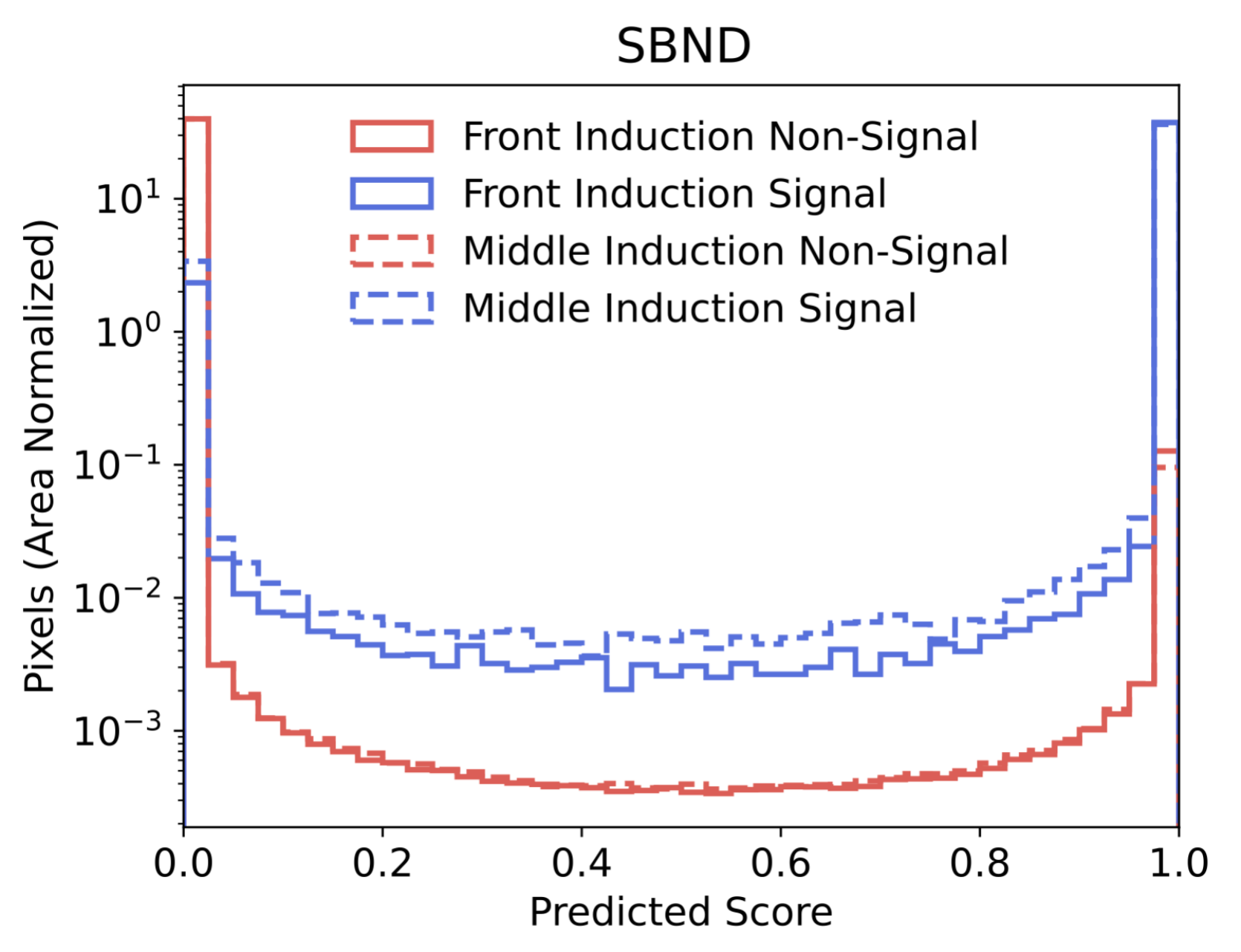}
    \caption{The DNN prediction score for true signal and non-signal pixels, evaluated for SBND front and middle induction planes. Shown for a sample of simulated BNB neutrinos with cosmic-ray activity included.}
    \label{fig:score_dist}
\end{figure}

Figure \ref{fig:score_dist} shows the distribution of DNN-predicted scores for the induction planes in SBND. The scores for true signal and non-signal pixels are well separated, both peaking near the edge bins. The default choice of threshold of 0.5 to classify pixels as regions of interest during signal processing ensures that the threshold lies well away from the regions of steep variation in the score distributions, ensuring stability and reliability in the classification.

\begin{figure}[t]
    \centering
    \includegraphics[width=0.45\textwidth]{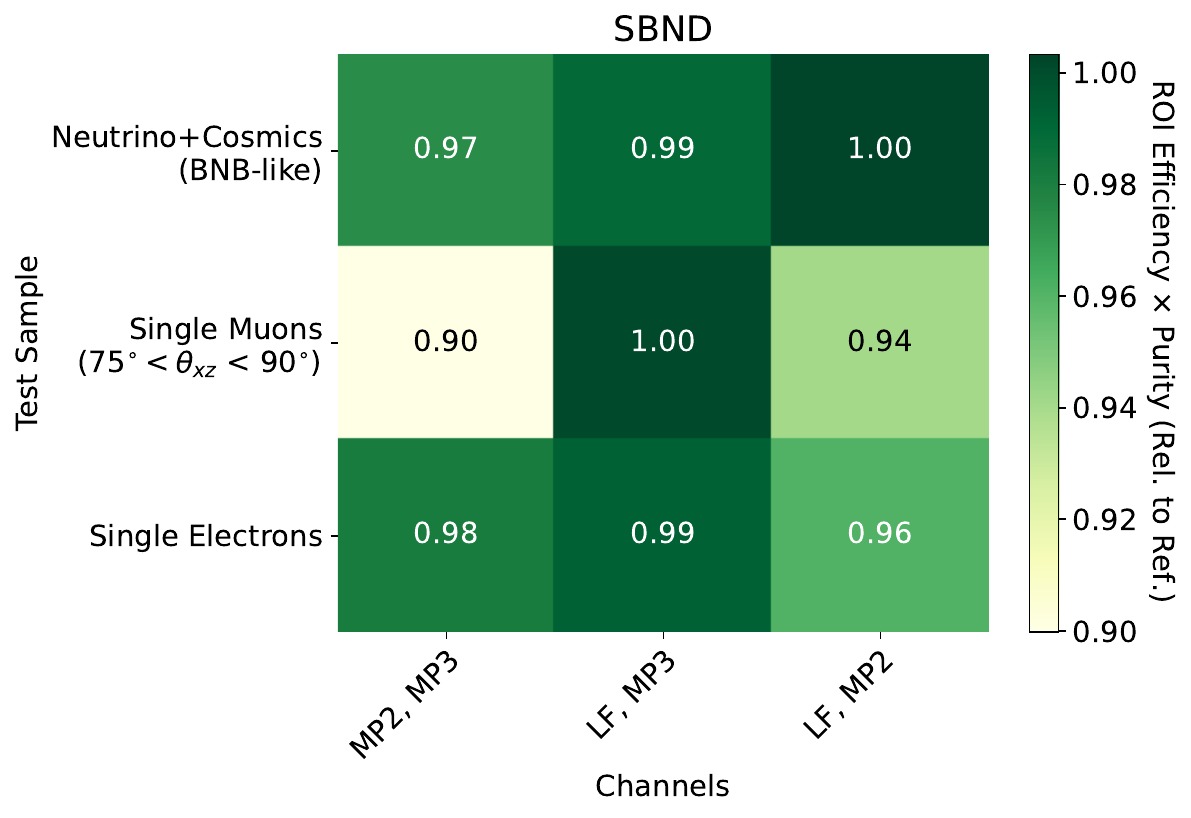}
    \caption{Comparison of ROI efficiency$\times$purity on different samples evaluated with networks trained with different combinations of input channels. The comparison leverages a sample of simulated BNB neutrinos with cosmic-ray activity included. The performance is shown as a relative scale to the reference performance from a network trained using all three input channels. The labels for each frame are introduced in Sec.~\ref{sec:pre-process}.}
    \label{fig:sbnd-inputchannel-effect}
\end{figure}

To assess the contribution of each input channel, we conducted ablation studies in which the DNN is trained using only two out of the three input channels (see Sec.~\ref{sec:pre-process} for their description). Figure \ref{fig:sbnd-inputchannel-effect} shows the performance of each network variant on three different types of test samples. These include BNB-like neutrino events with cosmics, single muons at prolonged angles, and single electron showers, in relative scale to the nominal case with all three input channels. We observe that MP2 has the least impact on the overall performance when excluded, with all samples losing less than 1\% of performance. Additionally, we find that the LF channel (the image obtained by applying the ROI filter to the waveforms) is particularly important for the prolonged tracks, whereas MP3 has a larger influence on showers. The results show that while the network maintains good performance on the neutrino with cosmics sample even with missing channels, more challenging cases like prolonged tracks and showers experience noticeably degraded performance when input channels are removed.

\begin{figure}[]
    \centering
    \includegraphics[width=0.5\textwidth]{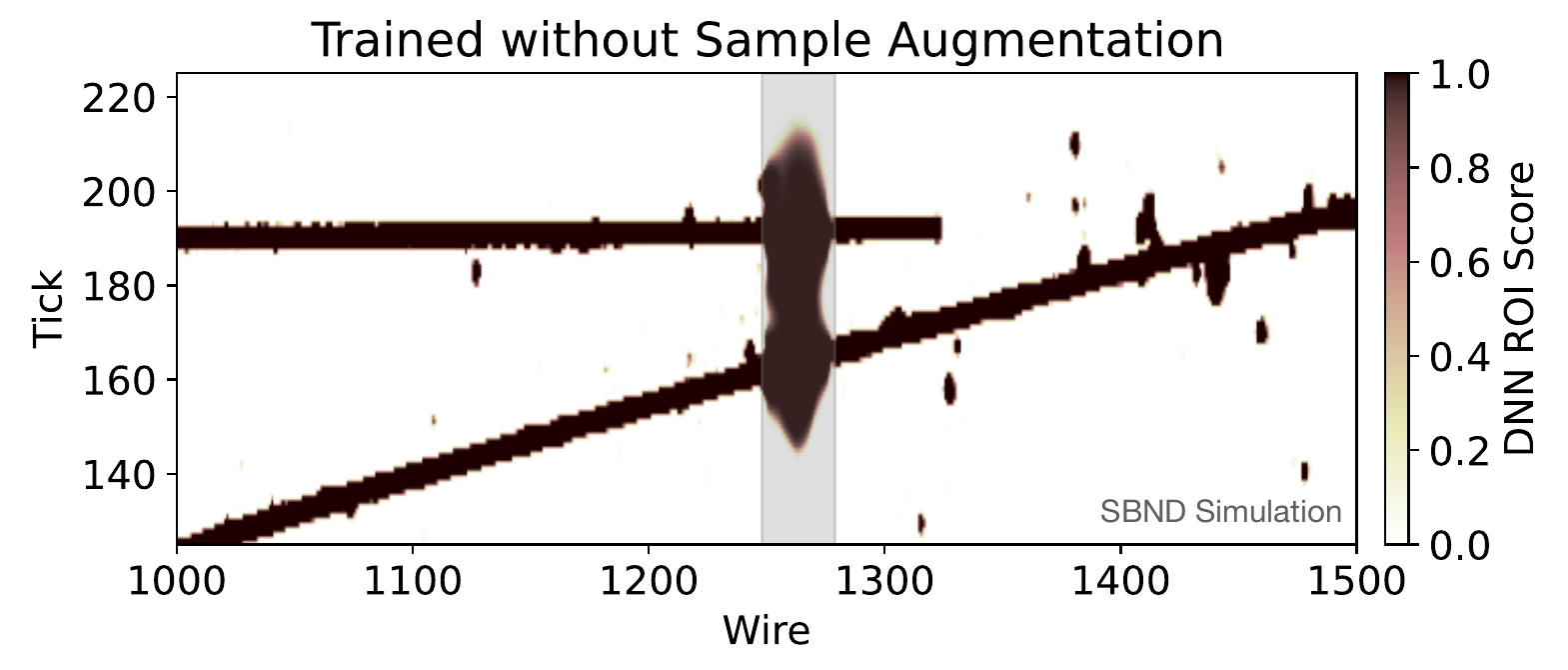}
    \includegraphics[width=0.5\textwidth]{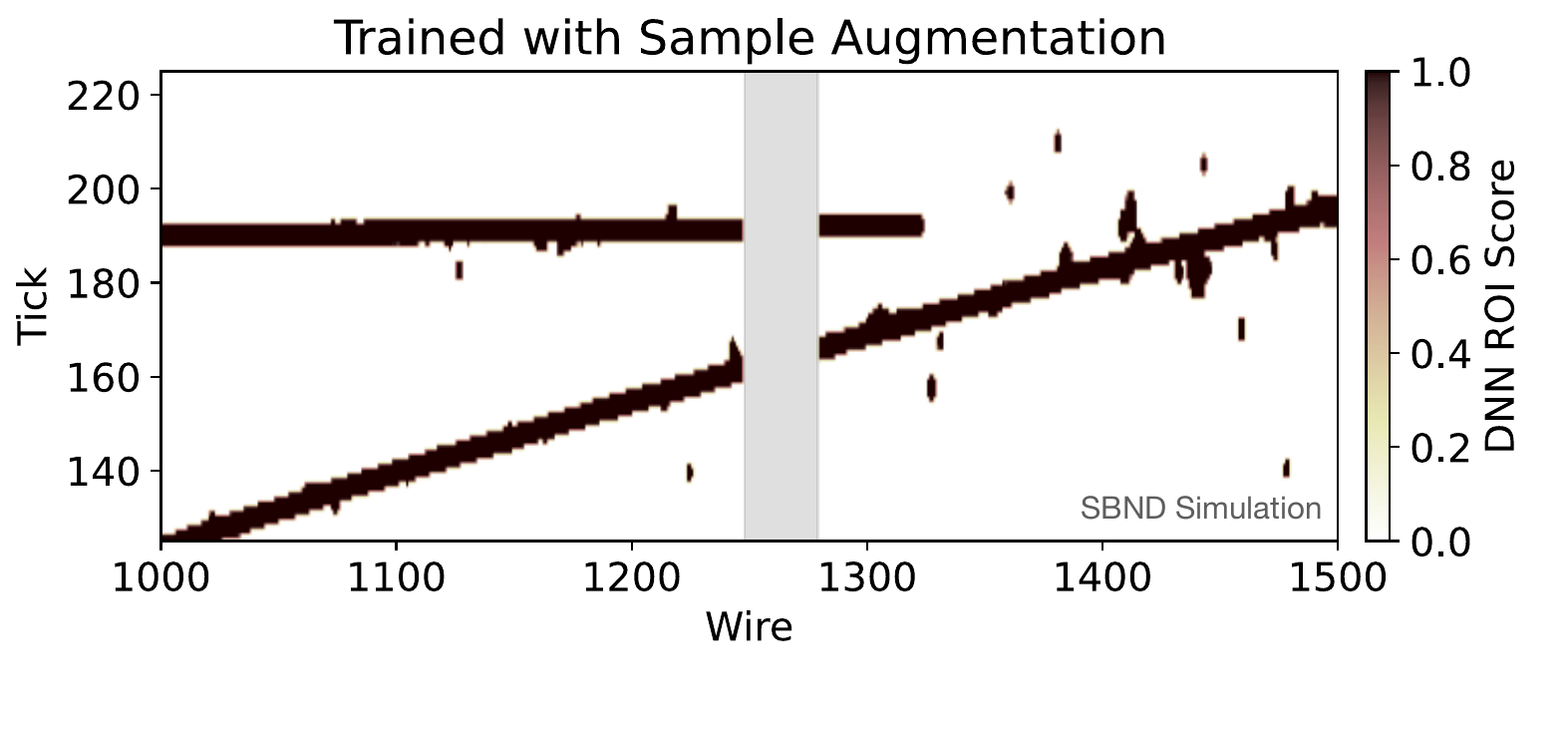}
    \caption{DNN ROI score prediction of a network trained on samples without augmentation (top) and with augmentation (bottom) on an example SBND MC event. The gray band marks the detector dead region.}
    \label{fig:sbnd-aug-effect}
\end{figure}

Our results in ICARUS and SBND indicate that DNN ROI is robust against detector performance variations, even when the training sample is not augmented to include them. However, we have found that augmentation is essential to remain robust against data corruption type effects. For example, Fig. \ref{fig:sbnd-aug-effect} shows the DNN score predictions on a sample event with dead wires using networks trained with and without sample augmentation. Results show that the network trained on augmented samples successfully learns to identify and ignore dead wires, while the network trained only on nominal samples predicts unphysical ROIs in those areas, as expected. The prediction on areas outside of the dead wire region remain reasonable and largely unaffected, even in the case of the network trained on samples without augmentation.

\begin{figure}[t]
    \centering
    \includegraphics[width=0.5\textwidth]{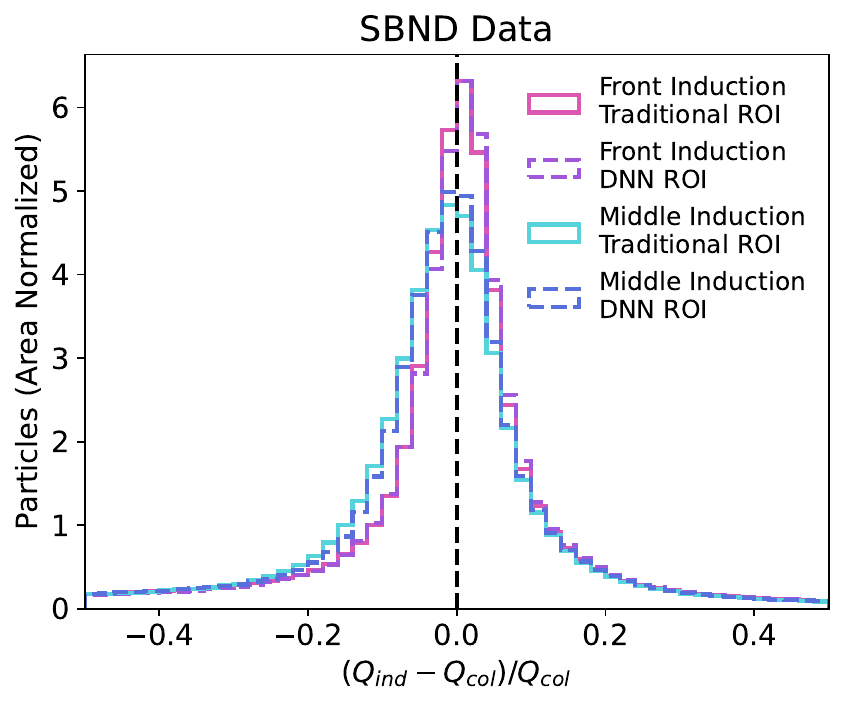}
    \caption{The fractional offset between the total measured charge of reconstructed particles on the induction planes relative to the collection plane for SBND cosmic data.}
    \label{fig:sbnd-chargeconsistency}
\end{figure}

To evaluate the performance and robustness of ROI finding on real data, we cannot use the truth information. Instead, we can assess the consistency of the reconstructed charge across the three wire planes. Due to the design of the detector, the induced and collected charge on the three wire planes is expected to be consistent for the same physical ionization. Figure \ref{fig:sbnd-chargeconsistency} shows a comparison of the extracted charge across the three planes for cosmic data at SBND. We compare the total charge of reconstructed particle objects that are matched across all three planes and have trajectories longer than \SI{10}{cm}. 

The charge ratio distributions show no significant peak bias between planes, indicating that both the DNN ROI and traditional methods preserve the expected inter-plane charge balance after calibration. The comparable peak value and width between traditional and DNN ROI demonstrate that DNN ROI achieves robust charge reconstruction at a level comparable to the traditional method in real detector data, providing validation of its applicability beyond simulation.

Finally, further optimization could extend the reach of the algorithm to include lower-energy depositions. The current charge threshold is primarily determined by two factors: the minimum charge required for a pixel to be considered when constructing the cross-plane coincidence frames (approximately three times the noise RMS for the plane of interest in three-plane matching and five times the noise RMS for others) and the charge threshold applied to define true ROI pixels in the target frame during network training. These thresholds effectively set the lower bound on what the model can learn to identify as ROI. These thresholds could be further tuned to target lower energy depositions, beyond the track-like and shower-like energy depositions from neutrino interactions mainly considered in this study. In addition, including radiogenic and other low-energy signatures in the training sample would also enable the algorithm to better identify them.

%% file: Conclusion.tex
\label{sec:conclusion}
We developed a DNN–based approach for ROI identification in LArTPCs of the SBN Program. The approach uses U-ResNet models, optimized for each detector and wire plane, that uses filtered waveforms and cross-plane matched masks as the input channels. The training dataset was carefully constructed to capture the ionization signal patterns of interest and detector performance variations observed in SBND and ICARUS.

By leveraging the full two-dimensional detector readout and cross-plane matching, DNN ROI overcomes the limitations of traditional wire-by-wire thresholding. It shows significant improvements across both low-level metrics, like pixel- and ROI-level efficiency and purity, and high-level metrics, like ionization charge extraction and particle energy reconstruction. The method also exhibits robustness against realistic variations in detector performance, like higher noise levels, wire plane intransparency, and argon impurities. Even in extreme cases, such as unresponsive wire plane regions, any resulting misidentification remains localized to the affected region.

DNN ROI is now used for SBND and ICARUS to maximize the experiments' physics capabilities. This method provides a flexible framework that can be quickly readapted to changing detector conditions. In the future, incorporating ML computing like GPUs will allow this approach to be scaled to future large-scale experiments, such as DUNE, ensuring broad applicability for LArTPCs.

%% file: acknowledgements.tex
This document was prepared by the ICARUS and SBND collaborations using the resources of the Fermi National Accelerator Laboratory (Fermilab), a U.S. Department of Energy, Office of Science, HEP User Facility. Fermilab is managed by FermiForward Discovery Group, LLC, acting under Contract No. 89243024CSC000002.

This work was also supported by Istituto Nazionale di Fisica Nucleare (INFN, Italy), EU Horizon 2020 Research and Innovation Program under the Marie Sklodowska-Curie Grant Agreement Nos. 822185, 858199, and101003460 and Horizon Europe Program research and innovation programme under the Marie Sklodowska-Curie Grant Agreement No. 101081478, the research contract per Law 240/2010, Art. 24 (3)(a), and D.G.R. 693/2023 (REF. PA:2023-20090/RER—CUP:J19J23000730002) by FSE+ 2021–2027. Furthermore the support of CERN in the detector overhauling within the Neutrino Platform framework, in the detector installation and commissioning, is acknowledged. This work was also supported by the Anusandhan National Research Foundation (ANRF, India) under the Ramanujan Fellowship (Grant No. RJF/2025/000203).

The ICARUS Collaboration would like to thank the MINOS Collaboration for having provided the Side CRT panels as well as Double Chooz (University of Chicago) for the Bottom CRT panels. We also acknowledge the contribution of many SBND colleagues, in particular for the development of a number of simulation, reconstruction and analysis tools which are shared within the SBN program.

The SBND Collaboration acknowledges the generous support of the following organizations: the U.S. Department of Energy, Office of Science, Office of High Energy Physics; the U.S. National Science Foundation; Los Alamos National Laboratory for LDRD funding, the Science and Technology Facilities Council (STFC), part of United Kingdom Research and Innovation (UKRI), the UKRI Future Leaders Fellowship (grant number MR/V022407/1), and The Royal Society; the Swiss National Science Foundation; the Spanish Ministerio de Ciencia, Innovacíon y Universidades (MICIU/ AEI/ 10.13039/ 501100011033) under grants No PRE2019-090468, CNS2022-136022, RYC2022-036471-I, PID2023-147949NB-C51 \& C53 and Comunidad
de Madrid (PEJ-2023-AI/COM-28399); the European Union’s Horizon 2020 research and innovation program under GA no 101004761 and the Marie Sklodowska-Curie grant agreements No 822185, 101081478, and 101003460; the São Paulo Research Foundation 1098 (FAPESP), the National Council of Scientific and Technological Development (CNPq) and Ministry of Science, Technology \& Innovations-MCTI of Brazil; the Minas Gerais research foundation (FAPEMIG), grants APQ-00544-23 and APQ-01249-24; the Anusandhan National Research Foundation (ANRF, India) under the Ramanujan Fellowship (Grant No. RJF/2025/000203). An award of computer time was provided by the ASCR Leadership Computing Challenge (ALCC) program. This research used resources of the Argonne Leadership Computing Facility, which is a U.S. Department of Energy Office of Science User Facility operated under contract DE-AC02-06CH11357.

%% file: Software.tex
\section{Samples and Software Location}
\label{app:software}

\subsection{ICARUS}

The ICARUS samples were generated with a local installation of \lstinline{icaruscode} \lstinline{v09_84_00}. The following fcl files were used for the generation and Geant4 stages:
\begin{itemize}
    \item BNB $\nu\,+$ Cosmic:\\{ \lstinline{prodcorsika_genie_protononly_icarus_numi_volDetEnclosure_tpc.fcl}}
    \item NuMI $\nu\,+$ Cosmic: \\\lstinline{prodcorsika_bnb_genie_protononly_icarus.fcl}
    \item BNB $\nu_e$:\\\lstinline{simulation_genie_icarus_bnb_nue.fcl}
    \item Higgs Portal Scalar (HPS) $e^+e^-$ decays:\\
    \lstinline{dissH_corsikap_kdif_M050.fcl}
    \item Extended muons ($\theta_{xz}: 81^\circ$)\\
    \lstinline{muons_81.fcl}
    \item Extended muons ($\theta_{xz}: 84^\circ$)\\
    \lstinline{muons_84.fcl}
    \item Extended muons ($\theta_{xz}: 86^\circ$)\\
    \lstinline{muons_86.fcl}
    \item Geant4:\\
    \lstinline{cosmics_g4_icarus.fcl}
\end{itemize}

The extended muons samples were each generated with a custom fcl configuration that inheirted from the sss fcl file in icaruscode. Here is an example for the \lstinline{muons_81.fcl}:
\begin{lstlisting}
'#include "prodsingle_common_icarus.fcl"

process_name: SinglesGen

outputs.out1.fileName: "prod_muon_3GeV_81deg_icarus_%tc_gen.root"

physics.producers.generator.PadOutVectors: true

physics.producers.generator.PDG: [13, 13, 13, 13, 13, 13]
physics.producers.generator.P0: [3] # GeV

physics.producers.generator.X0: [50,50,50,-50,-50,-50] # cm ?
physics.producers.generator.Y0: [-125, -50, 25, -125, -50, 25]
physics.producers.generator.Z0:[50,250,400, -50, -250, -400]
physics.producers.generator.Theta0XZ: [81,-81,81,-81,81,-81]
physics.producers.generator.Theta0YZ: [0]

physics.producers.generator.T0: [0] # us
physics.producers.generator.SigmaX: [0]
physics.producers.generator.SigmaY: [0]
physics.producers.generator.SigmaZ: [0]
physics.producers.generator.SigmaT: [0]
physics.producers.generator.SigmaP: [0]
physics.producers.generator.PosDist: 0
physics.producers.generator.AngleDist: "uniform"
physics.producers.generator.SigmaThetaXZ: [1]
physics.producers.generator.SigmaThetaYZ: [0]
\end{lstlisting}
This configuration generates six cosmic muons spaced apart, each with a $\theta_{xz}$ value of $81^\circ$. Other angles can be generated by changing the \lstinline{physics.producers.generator.Theta0XZ} parameter.

The detector simulation configuration is generated file-by-file by the \lstinline{omnidetector} package: \lstinline{https://github.com/gputnam/omnidetector}. The base fcl file is \lstinline{detsim_2d_icarus_fitFR.fcl}. For \lstinline{icaruscode} \lstinline{v09_84_00}, this represents a uniform detector simulation (without the YZ-intransparency sim enabled) with simulated noise and fit signal shapes on the middle induction and collection planes, as is detailed in the original ICARUS calibration paper~\cite{ICARUSCalibration}. The following fcl parameters are used to change the detector simulation in the ways described in this note:
\begin{itemize}
    \item Front induction gain: \lstinline{physics.producers.daq.wcls_main.structs.gain0}
    \item Middle induction gain: \lstinline{physics.producers.daq.wcls_main.structs.gain1}
    \item Collection gain: \lstinline{physics.producers.daq.wcls_main.structs.gain2}

    \item Front induction electronics shaping time: \lstinline{physics.producers.daq.wcls_main.structs.shaping0}
    \item Middle induction electronics shaping time: \lstinline{physics.producers.daq.wcls_main.structs.shaping1}
    \item Collection electronics shaping time: \lstinline{physics.producers.daq.wcls_main.structs.shaping2}

    \item Coherent noise scale: \lstinline{physics.producers.daq.wcls_main.structs.coh_noise_scale}
    \item Incoherent noise scale: \lstinline{physics.producers.daq.wcls_main.structs.coh_noise_scale}

    \item Electron lifetime (same in both cryostats): \lstinline{physics.producers.daq.wcls_main.structs.lifetime}

    \item Middle induction signal shape (due to transparency variation): \lstinline{physics.producers.daq.wcls_main.params.files_fields}

    In this case, the parameter is set to \lstinline{icarus_fnal_fit_ks_P0nom_P1bin<I>.json.bz2}, where \lstinline{<I>} is the index of the signal shape from 0 to 15. The most intransparent (biggest up-beak) is at $I=0$, the average signal is at $I=7$, and the least intransparent is at $I=15$.

\end{itemize}

\subsection{SBND}
The SBND training and test samples were generated with a local installation of \lstinline{sbndcode} \lstinline{v10_04_01}. The following fcl files were used for the generation and Geant4 stages:

\begin{itemize}
    \item BNB $\nu\,+$ Cosmic:\\{ \lstinline{prodoverlay_corsika_cosmics_proton_genie_rockbox_sce.fcl}\\
    \lstinline{standard_g4_sbnd.fcl}}
    \item BNB intrinsic $\nu_e$:{\\
    \lstinline{prodoverlay_corsika_cosmics_proton_genie_intrnue_spill_tpc_sbnd.fcl}\\
    \lstinline{g4_sce_lite.fcl}}
    \item Extended muons for different $\theta_{xz}$ ranges:\\
    \lstinline{theta_<th_lo>_<th_hi>_muons.fcl}:
    \begin{lstlisting}
        #include "prodsingle_sbnd_proj.fcl"

        physics.producers.generator.PadOutVectors: true
        physics.producers.generator.PDG: [13, 13, 13, 13, 13, 13]
        physics.producers.generator.P0: [3] # GeV
        physics.producers.generator.X0: [50,50,50,-50,-50,-50] 
        physics.producers.generator.Y0: [0]
        physics.producers.generator.Z0:[50,250,400,50,250,400]
        physics.producers.generator.Theta0XZ: [(th_lo+th_hi)/2.,-(th_lo+th_hi)/2.,
        (th_lo+th_hi)/2.,-(th_lo+th_hi)/2.,(th_lo+th_hi)/2.,-(th_lo+th_hi)/2.]
        physics.producers.generator.Theta0YZ: [0]
        physics.producers.generator.T0: [0] # us 
        physics.producers.generator.SigmaX: [0]
        physics.producers.generator.SigmaY: [0]
        physics.producers.generator.SigmaZ: [0]
        physics.producers.generator.SigmaT: [0]
        physics.producers.generator.SigmaP: [0]
        physics.producers.generator.PosDist: 0
        physics.producers.generator.AngleDist: "uniform"
        physics.producers.generator.SigmaThetaXZ: [(th_hi-th_lo)/2.]
        physics.producers.generator.SigmaThetaYZ: [0]
    \end{lstlisting}
    
    \lstinline{g4_sce_lite.fcl}
    \item Single electrons for different energy ranges:\\
    \lstinline{electron_shower_<E_lo>_<E_hi>.fcl}
    \begin{lstlisting}
        #include "prodsingle_sbnd_proj.fcl"

        physics.producers.generator.PadOutVectors: true
        physics.producers.generator.PDG: [11,11]
        physics.producers.generator.PosDist: 0   #Flat position dist.
        physics.producers.generator.X0: [-100,100 ]
        physics.producers.generator.SigmaX: [0] # set x position
        physics.producers.generator.Y0: [0]
        physics.producers.generator.SigmaY: [150 ]
        physics.producers.generator.Z0:[250 ]
        physics.producers.generator.SigmaZ: [200 ]
        physics.producers.generator.P0:[ (E_lo+E_hi)/2.]
        physics.producers.generator.SigmaP:[ (E_hi-E_lo)/2.]
        physics.producers.generator.Theta0XZ: [-20, 20]
        physics.producers.generator.Theta0YZ: [0]
        physics.producers.generator.SigmaThetaXZ: [10]
        physics.producers.generator.SigmaThetaYZ: [0]
        
        services.NuRandomService.policy: "random"
    \end{lstlisting}
    \lstinline{g4_sce_lite.fcl}
\end{itemize}

All training was performed on Fermilab Elastic Analysis Facility (EAF), with NVIDIA A100 GPUs with memory 40GB with driver version 560.35.03. We used the PyTorch framework with python 3.10.18 cuda version 12.6.

%% file: ICARUSFilterOptimizaton.tex
\section{ICARUS Filter Parameter Optimization}
\label{app:icarus-filter-optimization}

\begin{figure}
    \centering
    \includegraphics[width=0.32\linewidth]{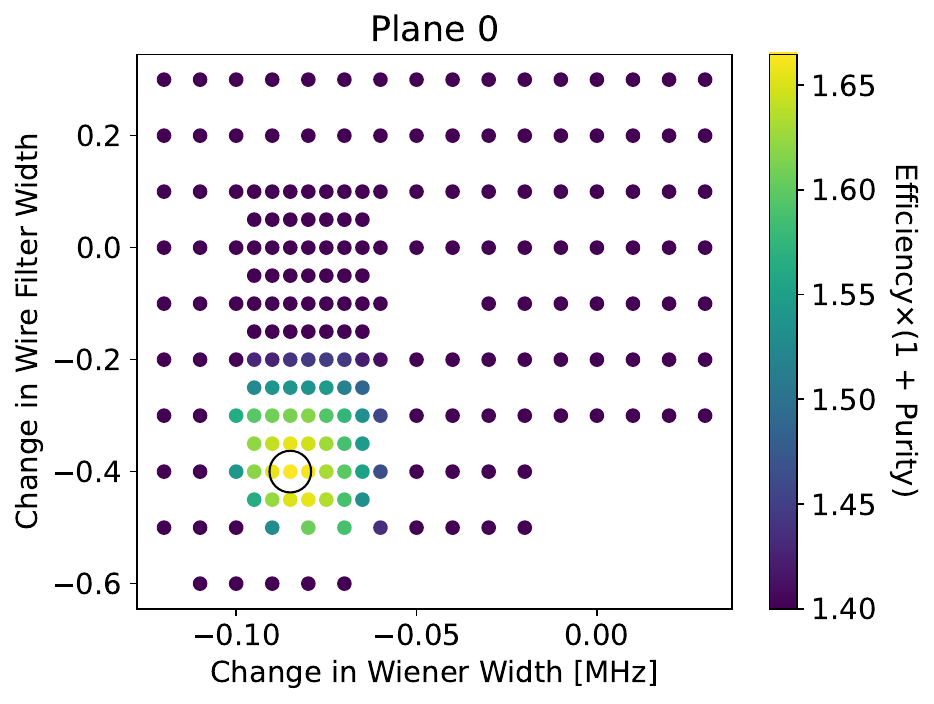}
    \includegraphics[width=0.32\linewidth]{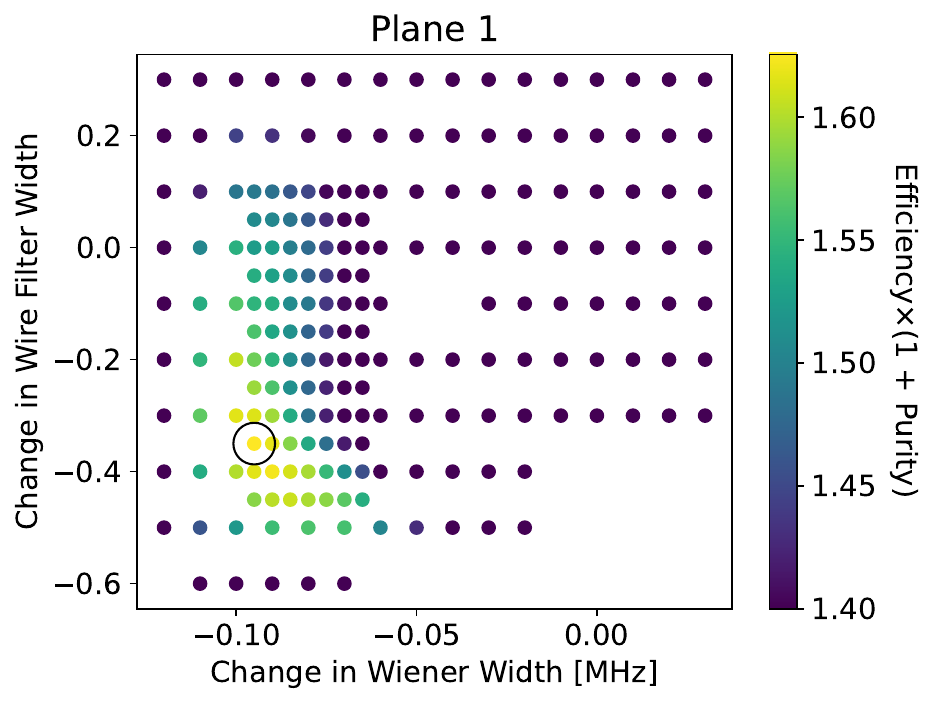}
    \includegraphics[width=0.32\linewidth]{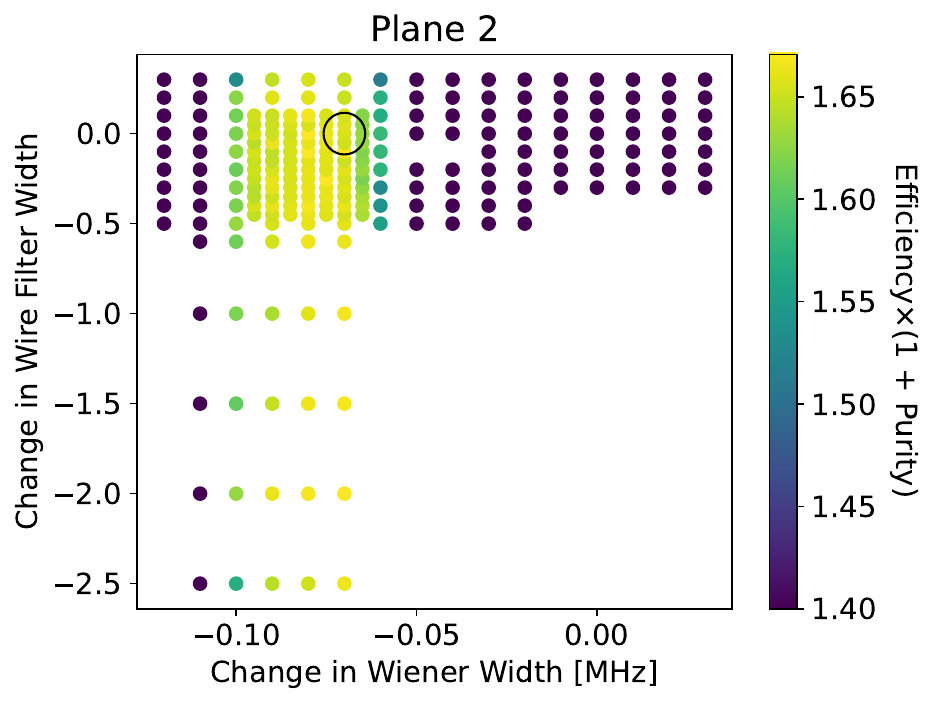}

    \includegraphics[width=0.32\linewidth]{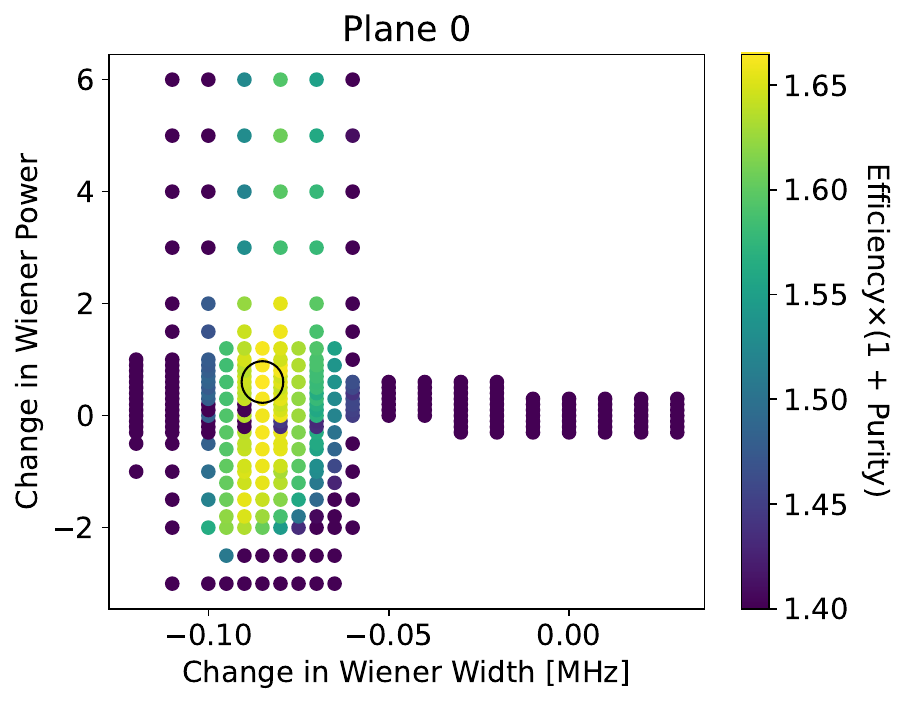}
    \includegraphics[width=0.32\linewidth]{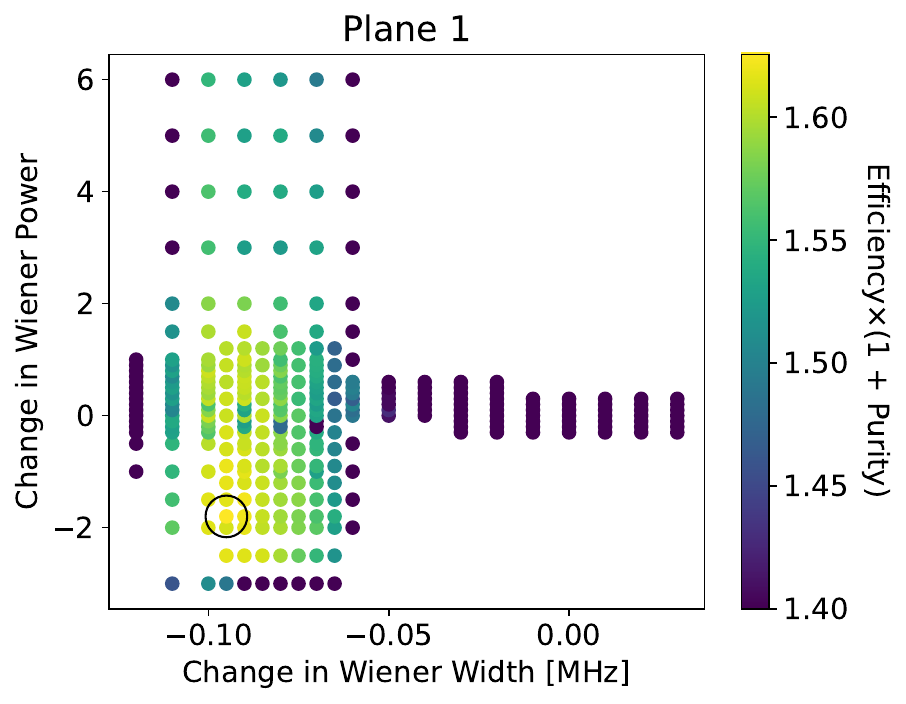}
    \includegraphics[width=0.32\linewidth]{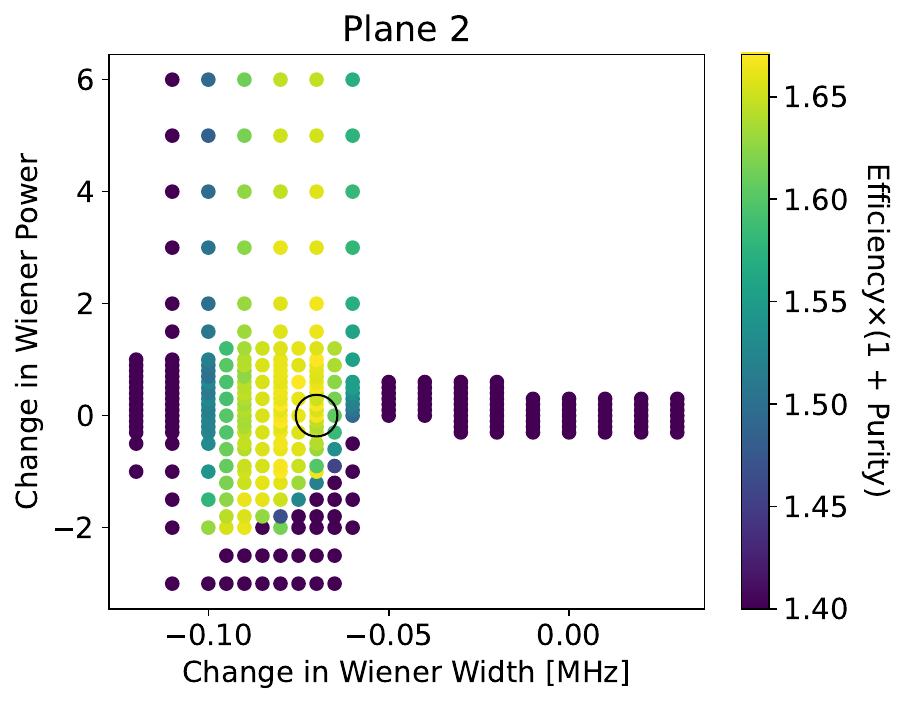}

    \includegraphics[width=0.32\linewidth]{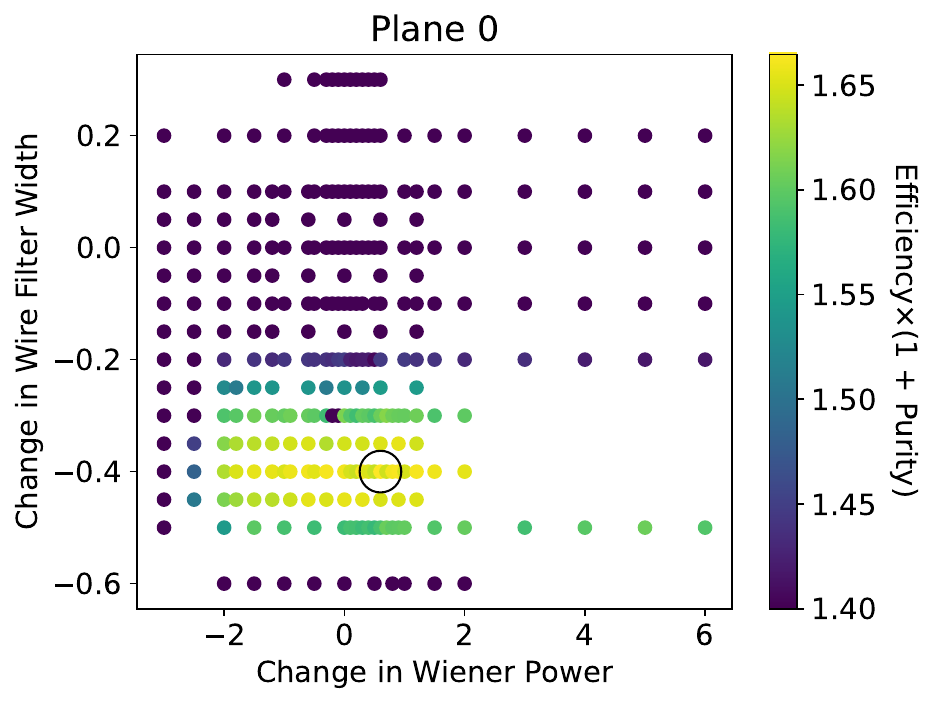}
    \includegraphics[width=0.32\linewidth]{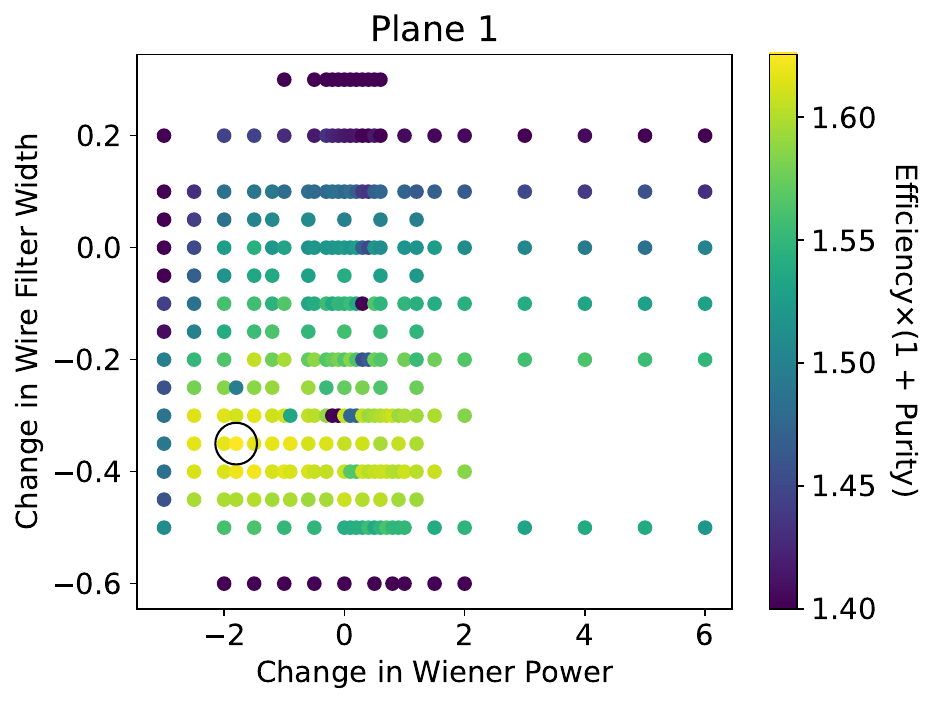}
    \includegraphics[width=0.32\linewidth]{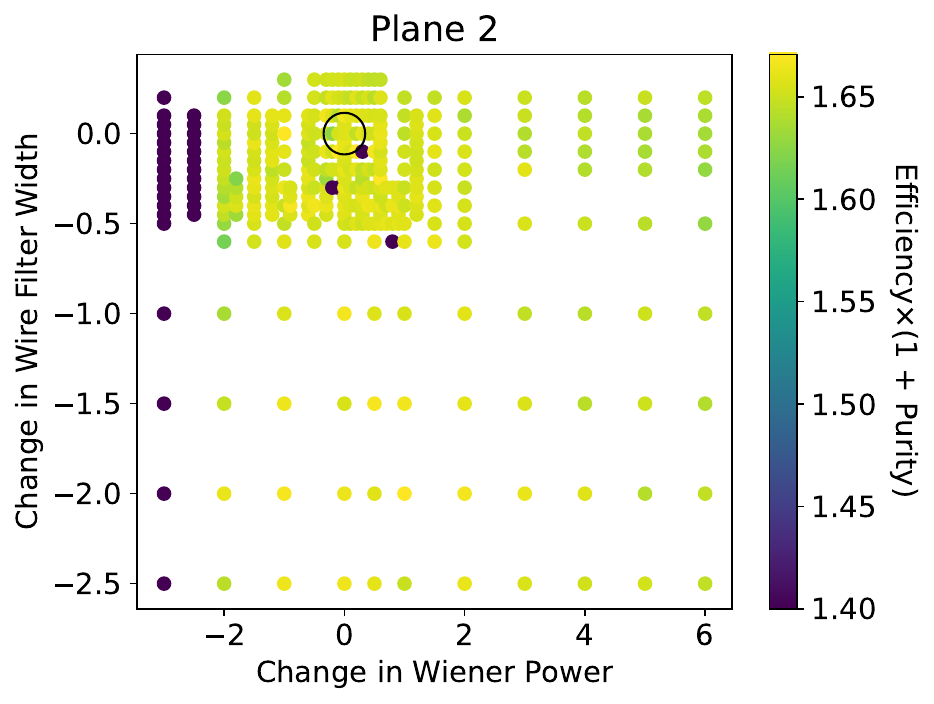}

    \caption{Result of optimization process for ROI and wire filters in Planes 0,1, and 2 (columns left to right). The 3D parameter scan is shown projected into 2D in all three pairings, one in each row.}
    \label{fig:filter-opt-2D}
\end{figure}

\begin{figure}
    \centering
    \includegraphics[width=0.32\linewidth]{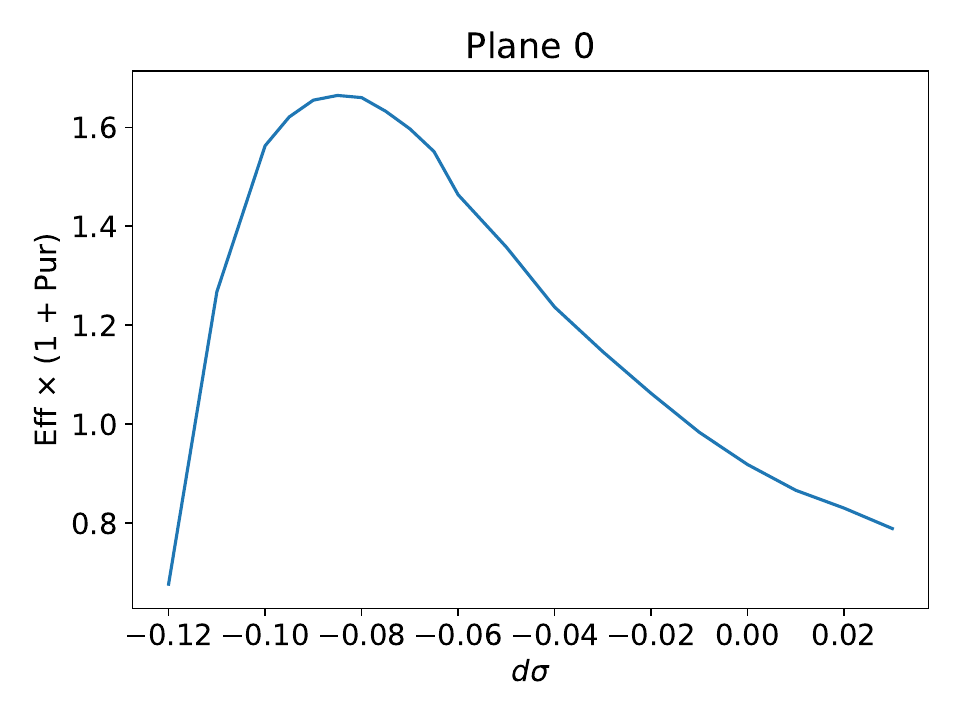}
    \includegraphics[width=0.32\linewidth]{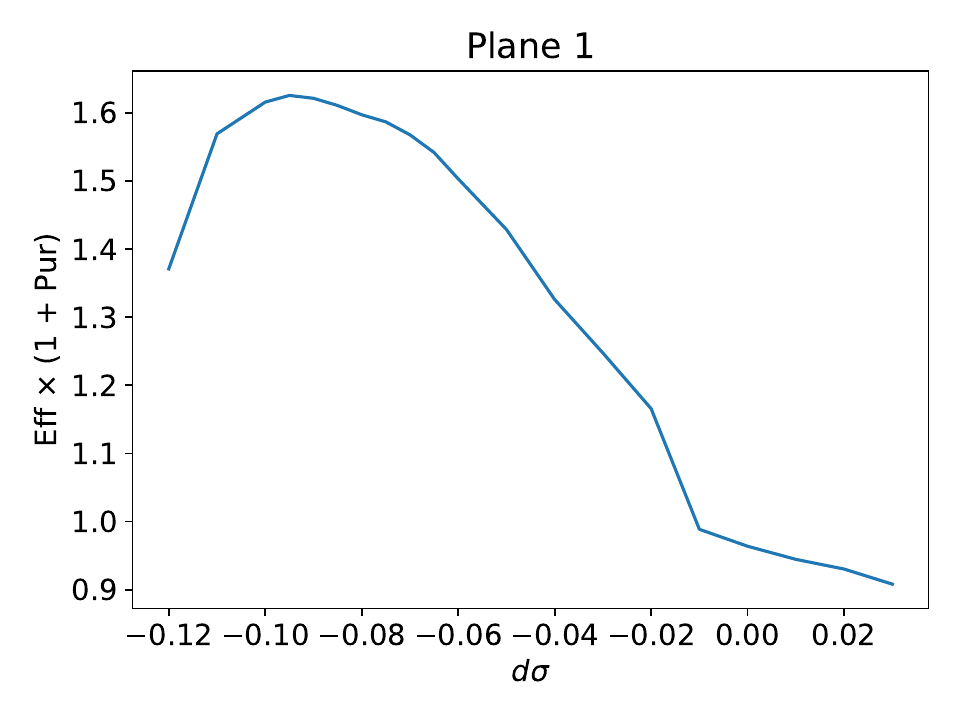}
    \includegraphics[width=0.32\linewidth]{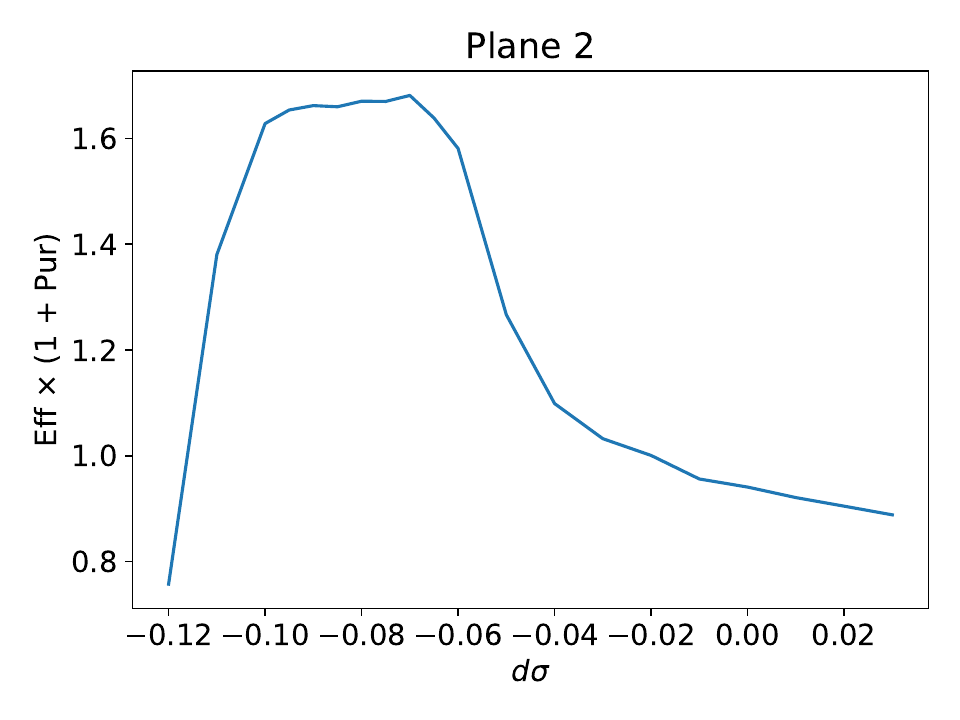}

    \includegraphics[width=0.32\linewidth]{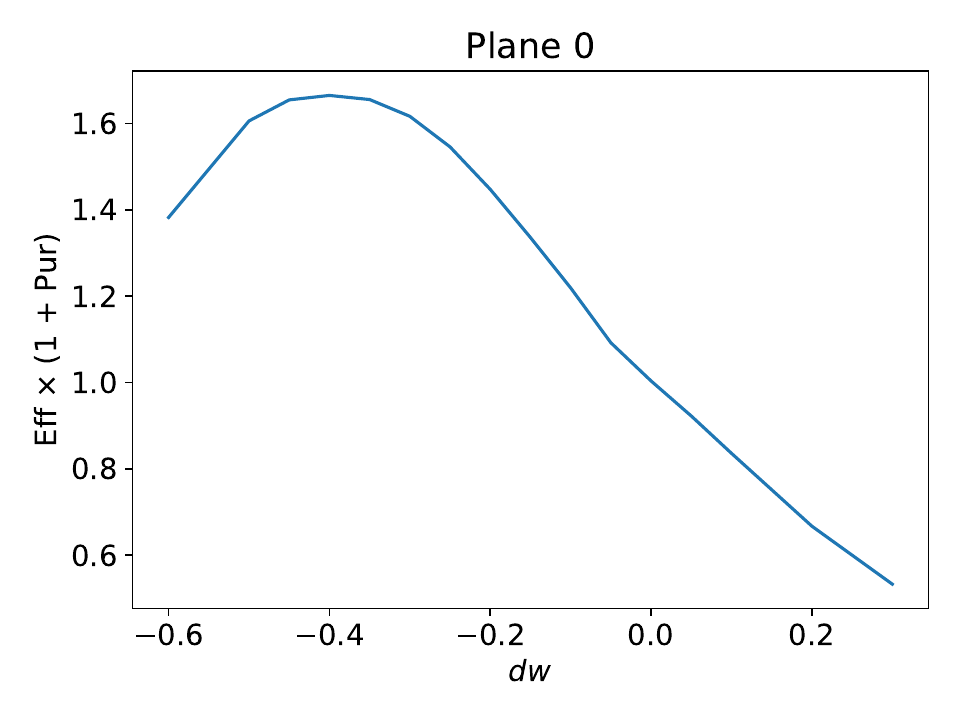}
    \includegraphics[width=0.32\linewidth]{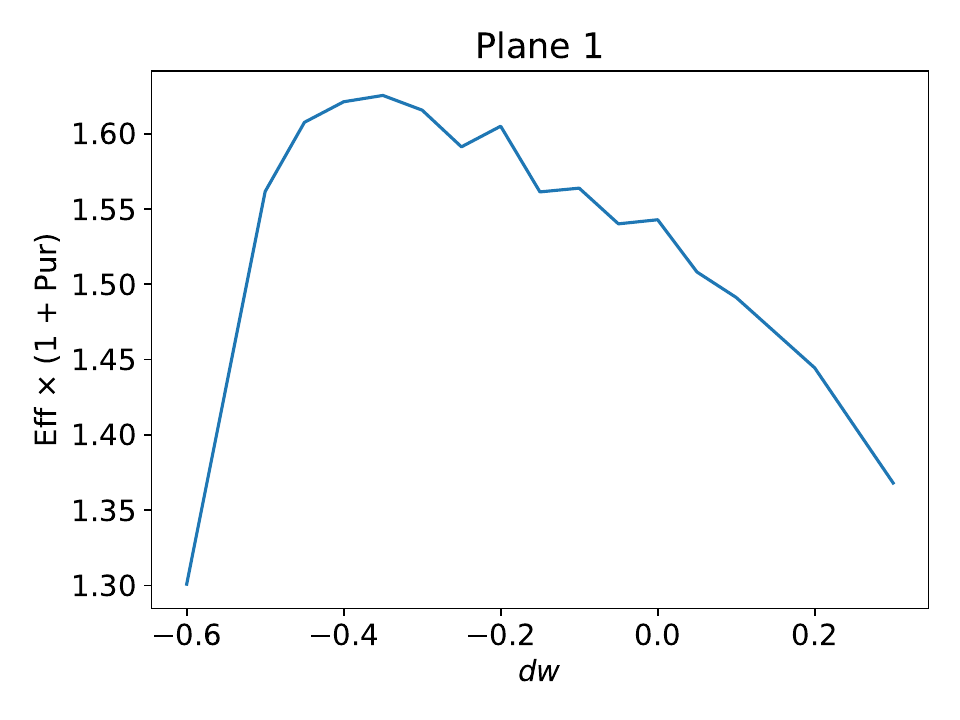}
    \includegraphics[width=0.32\linewidth]{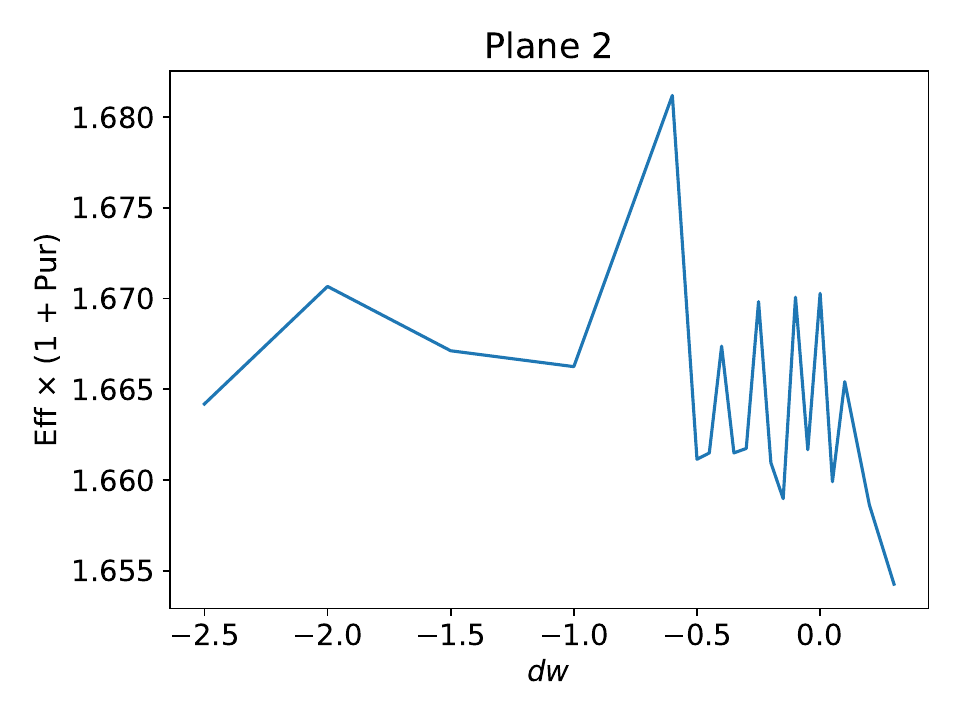}

    \includegraphics[width=0.32\linewidth]{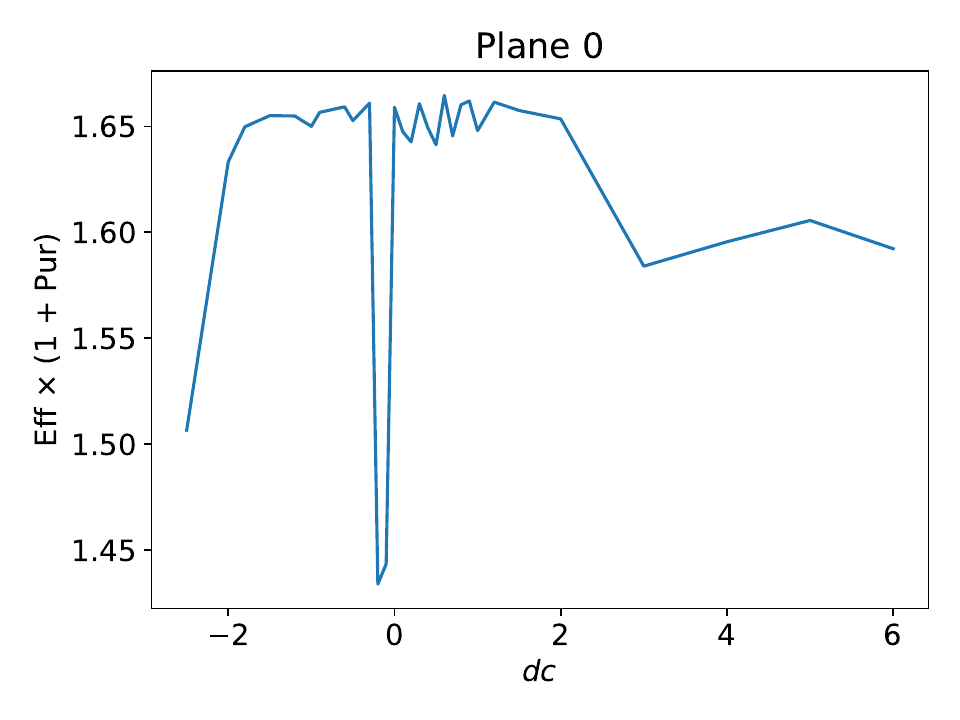}
    \includegraphics[width=0.32\linewidth]{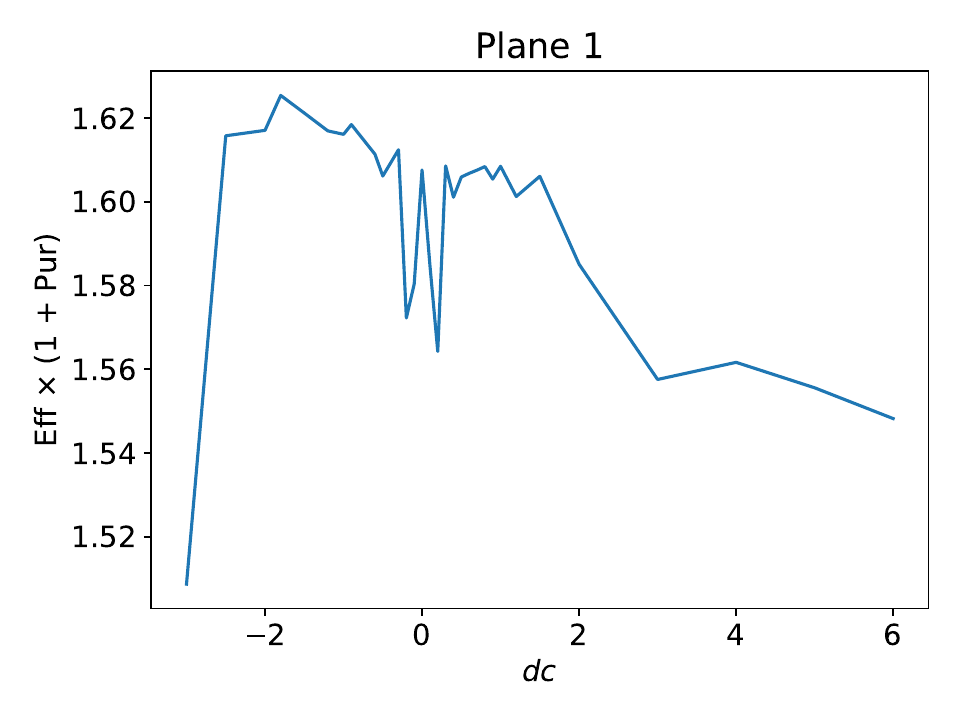}
    \includegraphics[width=0.32\linewidth]{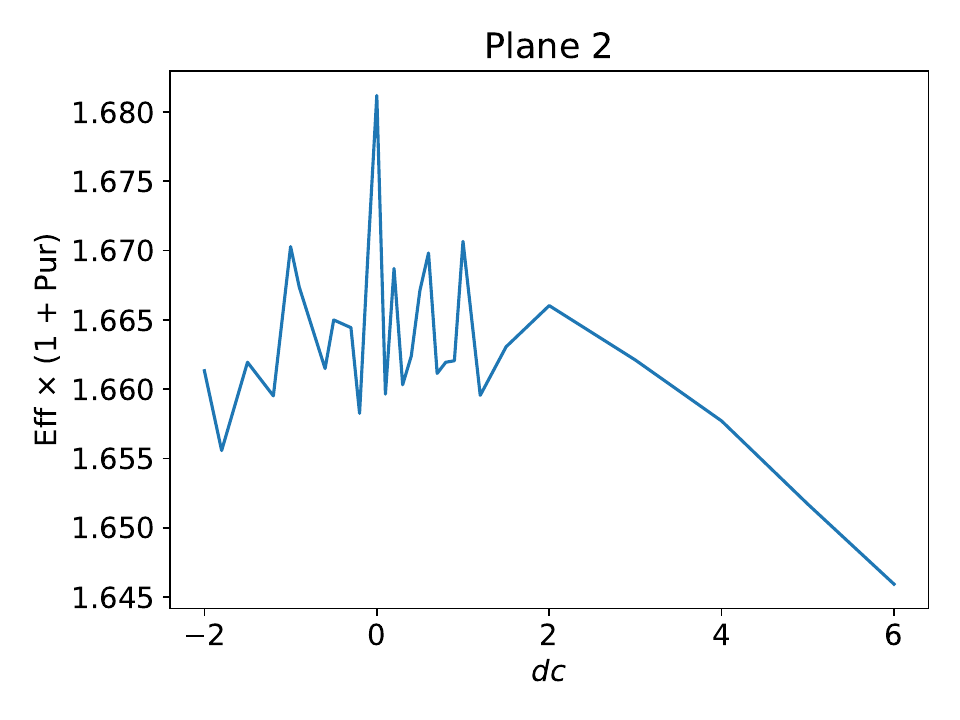}

    \caption{Result of optimization process for ROI and wire filters in Planes 0,1, and 2 (columns left to right). The 3D parameter scan is shown projected into 1D in all three parameters, one in each row.}
    \label{fig:filter-opt-1D}
\end{figure}

\begin{figure}
    \centering
    \includegraphics[width=0.32\linewidth]{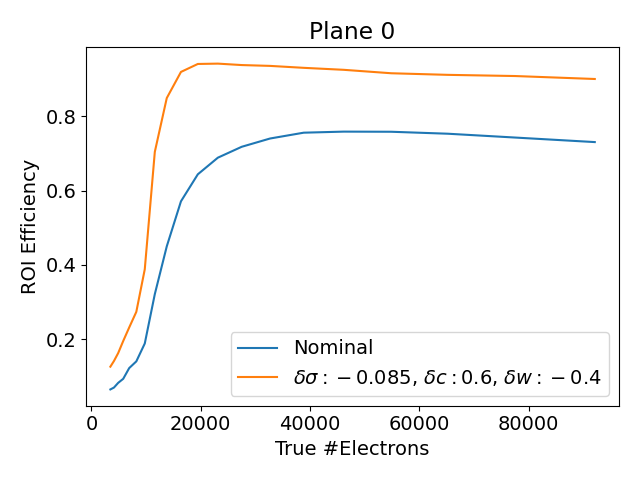}
    \includegraphics[width=0.32\linewidth]{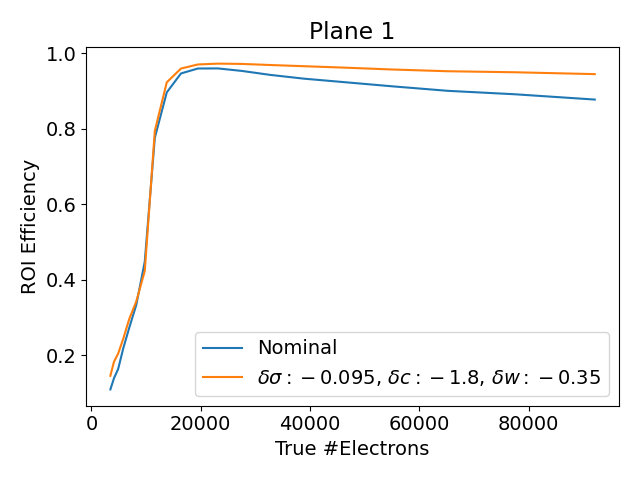}
    \includegraphics[width=0.32\linewidth]{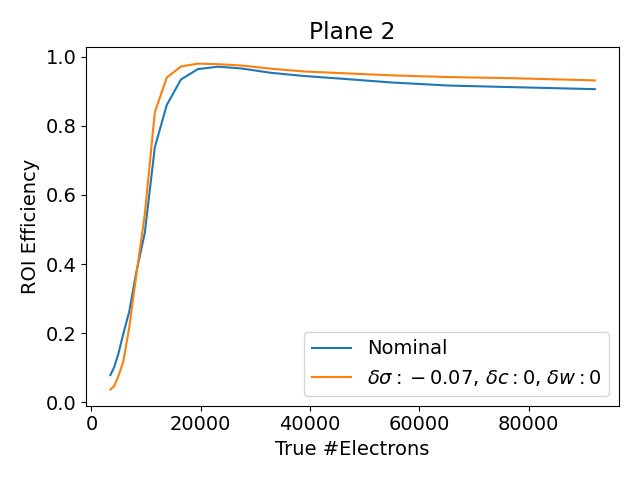}

    \includegraphics[width=0.32\linewidth]{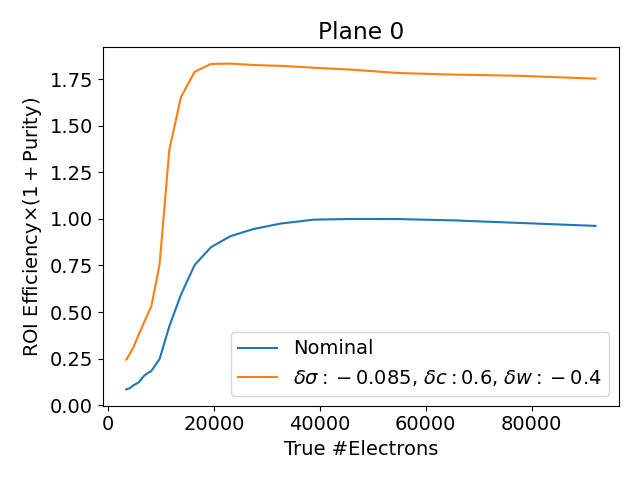}
    \includegraphics[width=0.32\linewidth]{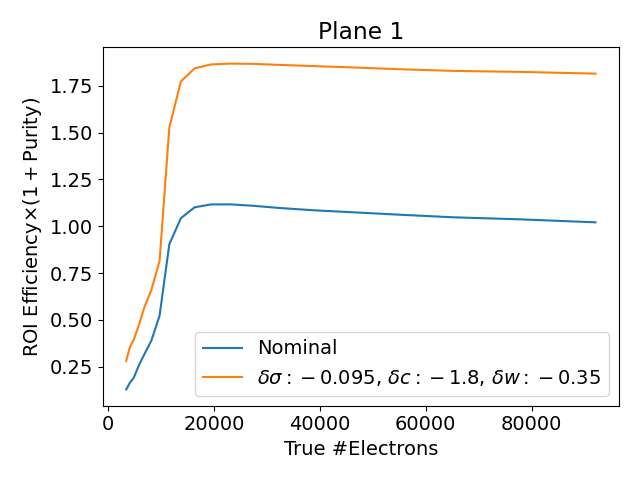}
    \includegraphics[width=0.32\linewidth]{filter-opt-figures/Plane2_opt_v_nominal_eff.png}
    \caption{ROI identification efficiency (top) and efficiecny $\times$ purity (bottom) on each plane (left to right). Shown for the initial filter parameters, compared to post-optimization.}
    \label{fig:filter-opt-comp}
\end{figure}

To optimize the filter values in ICARUS, we performed scans of the parameters in the filter functions, optimizing a chosen metric of the filter performance. First, we simultaneously optimized the wire filter (Eq.~\ref{eq:wirefilter}) and the ROI filter (Eq.~\ref{eq:ROIfilter}). The metric we optimized for was the ROI Efficiency$\times(1+$ ROI Purity$)$. The $1+$Purity factor de-weights the importance of the purity, relative to the efficiency, in the optimization. This is preferable because high-level clustering algorithms can remove noise hits, but nothing can be done to recover depositions that are ``missed'' by the ROI algorithm. Furthermore, we found that the optimization results for ROI Efficiency$\times(1+$ ROI Purity$)$ were consitent with that of ROI Efficiency$\times$ ROI Purity. The result of these scans are shown in figures~\ref{fig:filter-opt-2D} and ~\ref{fig:filter-opt-1D}, projected into 2D and 1D, respectively. The performance of the filter performance pre- and post-optimization is shown per-plane in figure~\ref{fig:filter-opt-comp}.

\begin{figure}
    \centering
    \includegraphics[width=0.32\linewidth]{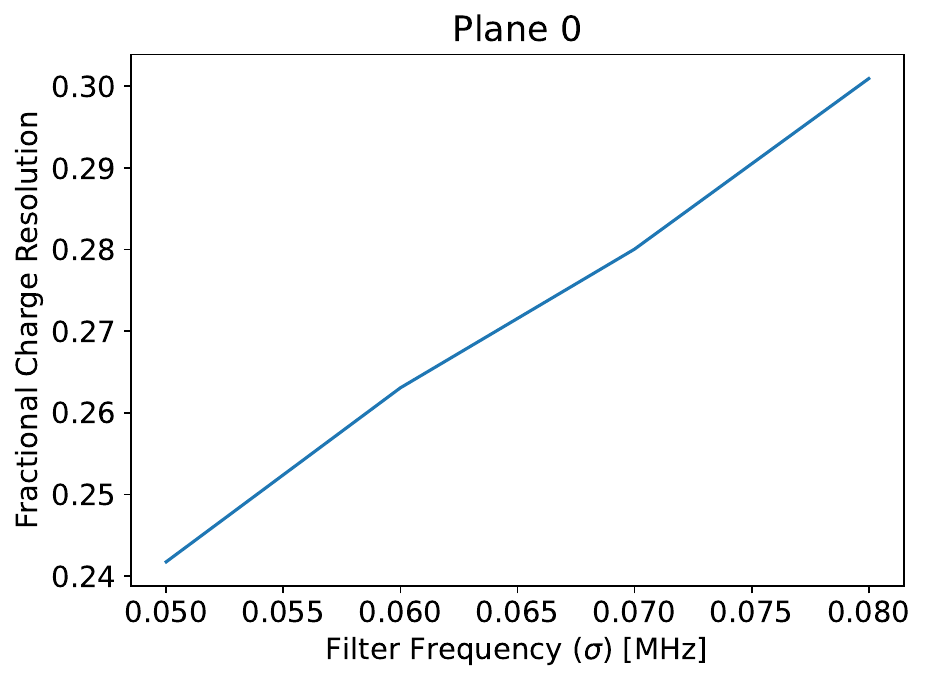}
    \includegraphics[width=0.32\linewidth]{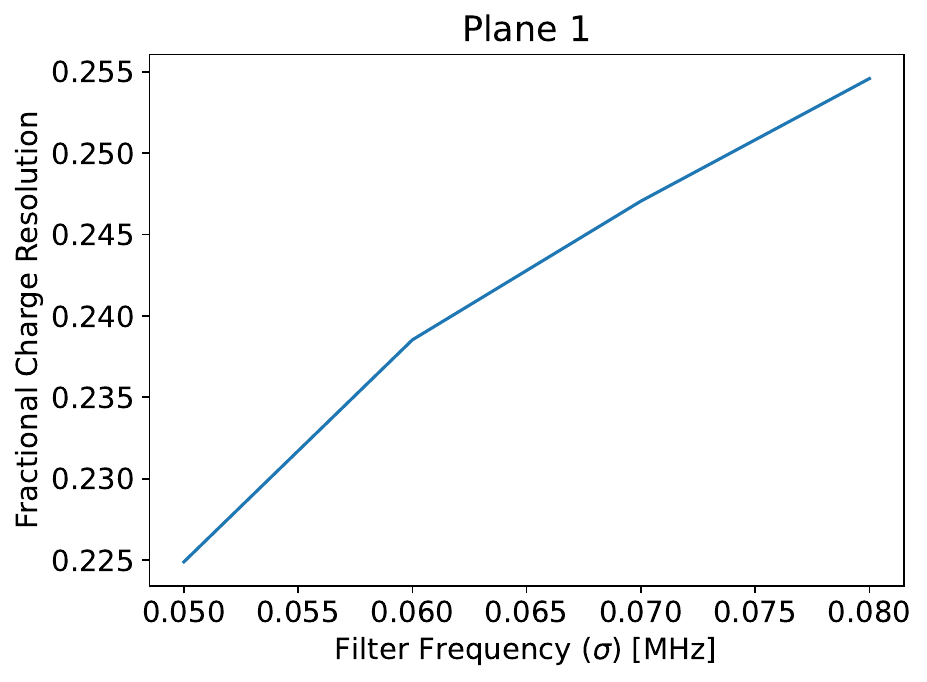}
    \includegraphics[width=0.32\linewidth]{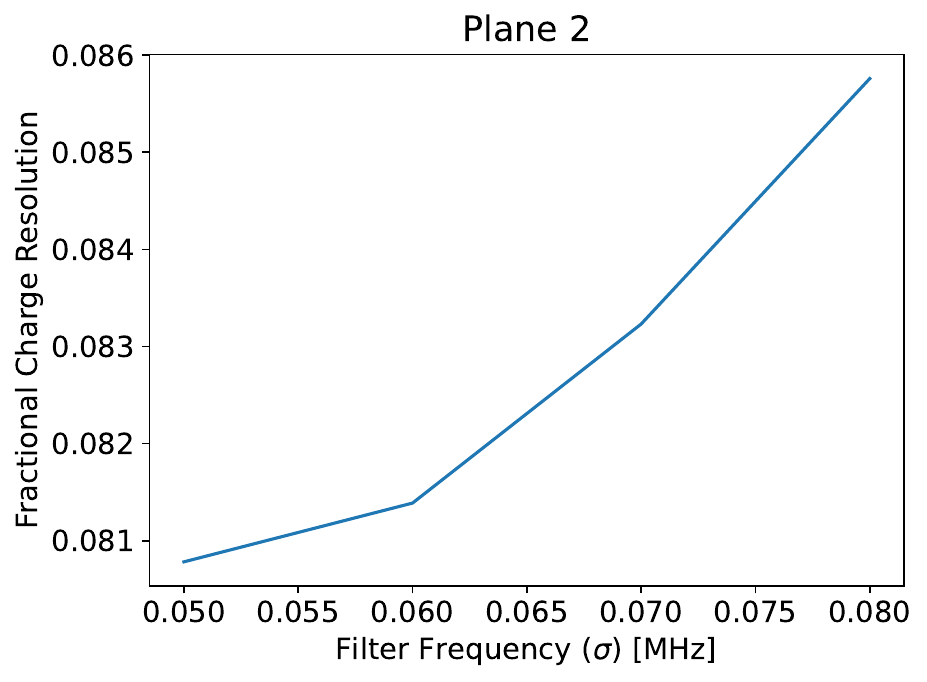}

    \includegraphics[width=0.32\linewidth]{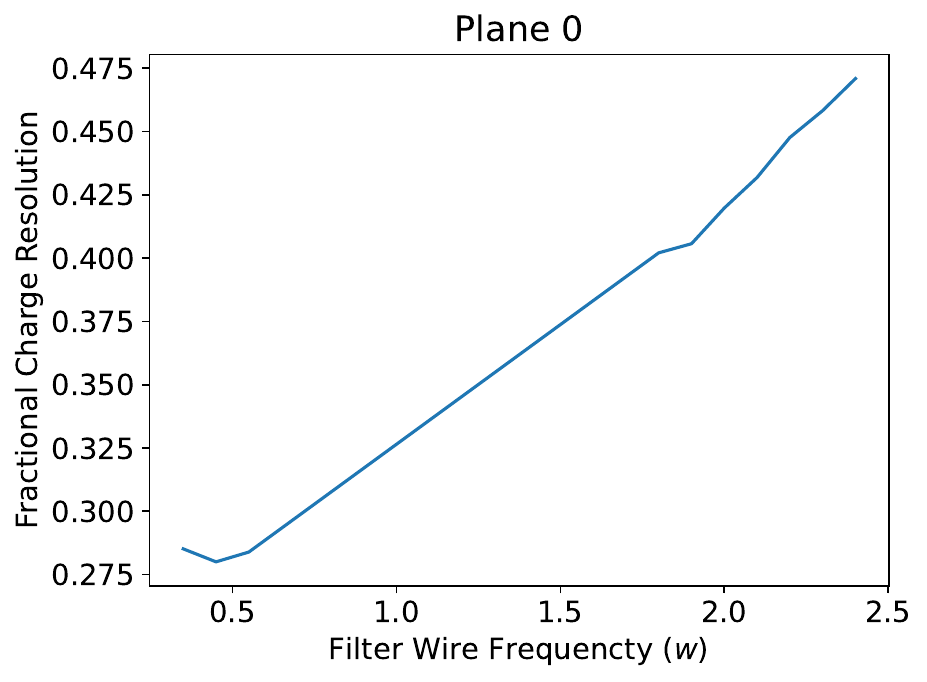}
    \includegraphics[width=0.32\linewidth]{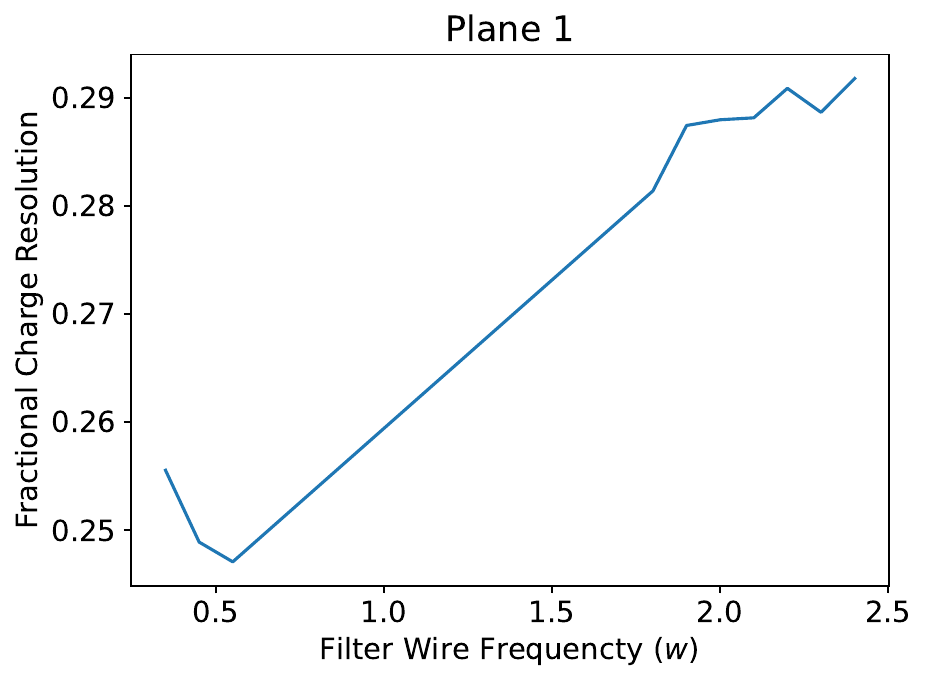}
    \includegraphics[width=0.32\linewidth]{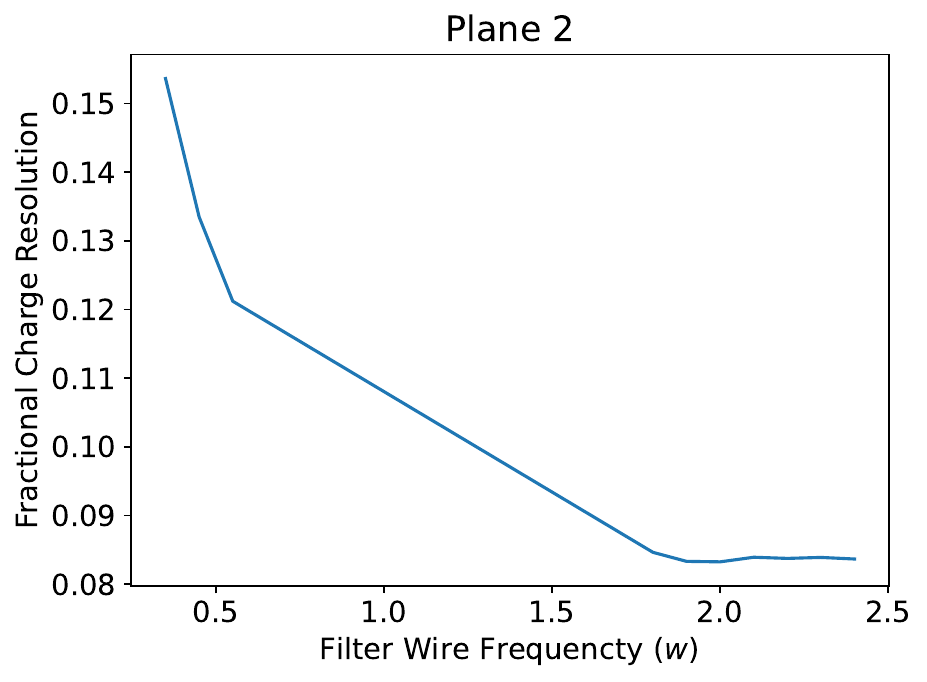}

    \caption{Charge resolution (as obtained by reconstructed hits) shown for varying charge filter and wire filter widths, projected into 1D. Shown for planes 0-2, left to right.}
    \label{fig:filter-charge-opt}
\end{figure}

After the ROI and wire filter optimization, we next optimized the charge extraction filter (Eq.~\ref{eq:Qfilter}), simultaneously with the wire filter width. The same wire filter width is applied in both the ROI identification and charge extraction steps, so this re-optimization was performed primarily to ensure that the two objectives gave similar results. In this case, the charge resolution of hit reconstruction is optimized. The result of this optimization is shown in figure~\ref{fig:filter-charge-opt}. Ultimately, each plane optimized for as small a charge filter width as possible (widest in the time domain). However, this result had to be balanced against the need to keep deconvolved wire signals narrow in the time domain (so as to not degrade spatial resolution) and consistent with the widths of the ROI filter output (so that ROI windows would be correctly applied). Balancing these considerations, we obtained the optimized filter values listed in table~\ref{tab:filters} in the main text.

%% file: NetworkOptimization.tex
\section{Network Optimization}
\label{app:network-optimization}

We explored 3 U-Net architecture variants: U-Net, U-ResNet, and Nested U-Net. A model for each architecture was optimized, and the best performing network was chosed. Optimization was performed through hyperparameter scan of learning rate, optimizer types, and optimizer parameters, and the batch size was chosen as the largest possible value permitted by the memory.

The network architecture model is available in \href{https://github.com/wjdanswjddl/Pytorch-UNet/blob/master/predict.py}{Pytorch-UNet} repository. The best performing network for ICARUS is a U-ResNet, summary for plane 0:
\begin{lstlisting}
    ----------------------------------------------------------------
        Layer (type)               Output Shape         Param #
================================================================
            Conv2d-1        [-1, 8, 1056, 1024]              32
       BatchNorm2d-2        [-1, 8, 1056, 1024]              16
            Conv2d-3        [-1, 8, 1056, 1024]             224
       BatchNorm2d-4        [-1, 8, 1056, 1024]              16
              ReLU-5        [-1, 8, 1056, 1024]               0
            Conv2d-6        [-1, 8, 1056, 1024]             584
            inconv-7        [-1, 8, 1056, 1024]               0
            Conv2d-8         [-1, 16, 528, 512]             144
       BatchNorm2d-9         [-1, 16, 528, 512]              32
      BatchNorm2d-10        [-1, 8, 1056, 1024]              16
             ReLU-11        [-1, 8, 1056, 1024]               0
           Conv2d-12         [-1, 16, 528, 512]           1,168
      BatchNorm2d-13         [-1, 16, 528, 512]              32
             ReLU-14         [-1, 16, 528, 512]               0
           Conv2d-15         [-1, 16, 528, 512]           2,320
   residual_block-16         [-1, 16, 528, 512]               0
             down-17         [-1, 16, 528, 512]               0
           Conv2d-18         [-1, 32, 264, 256]             544
      BatchNorm2d-19         [-1, 32, 264, 256]              64
      BatchNorm2d-20         [-1, 16, 528, 512]              32
             ReLU-21         [-1, 16, 528, 512]               0
           Conv2d-22         [-1, 32, 264, 256]           4,640
      BatchNorm2d-23         [-1, 32, 264, 256]              64
             ReLU-24         [-1, 32, 264, 256]               0
           Conv2d-25         [-1, 32, 264, 256]           9,248
   residual_block-26         [-1, 32, 264, 256]               0
             down-27         [-1, 32, 264, 256]               0
           Conv2d-28         [-1, 32, 132, 128]           1,056
      BatchNorm2d-29         [-1, 32, 132, 128]              64
      BatchNorm2d-30         [-1, 32, 264, 256]              64
             ReLU-31         [-1, 32, 264, 256]               0
           Conv2d-32         [-1, 32, 132, 128]           9,248
      BatchNorm2d-33         [-1, 32, 132, 128]              64
             ReLU-34         [-1, 32, 132, 128]               0
           Conv2d-35         [-1, 32, 132, 128]           9,248
   residual_block-36         [-1, 32, 132, 128]               0
             down-37         [-1, 32, 132, 128]               0
         Upsample-38         [-1, 32, 264, 256]               0
           Conv2d-39         [-1, 16, 264, 256]           1,040
      BatchNorm2d-40         [-1, 16, 264, 256]              32
      BatchNorm2d-41         [-1, 64, 264, 256]             128
             ReLU-42         [-1, 64, 264, 256]               0
           Conv2d-43         [-1, 16, 264, 256]           9,232
      BatchNorm2d-44         [-1, 16, 264, 256]              32
             ReLU-45         [-1, 16, 264, 256]               0
           Conv2d-46         [-1, 16, 264, 256]           2,320
   residual_block-47         [-1, 16, 264, 256]               0
               up-48         [-1, 16, 264, 256]               0
         Upsample-49         [-1, 16, 528, 512]               0
           Conv2d-50          [-1, 8, 528, 512]             264
      BatchNorm2d-51          [-1, 8, 528, 512]              16
      BatchNorm2d-52         [-1, 32, 528, 512]              64
             ReLU-53         [-1, 32, 528, 512]               0
           Conv2d-54          [-1, 8, 528, 512]           2,312
      BatchNorm2d-55          [-1, 8, 528, 512]              16
             ReLU-56          [-1, 8, 528, 512]               0
           Conv2d-57          [-1, 8, 528, 512]             584
   residual_block-58          [-1, 8, 528, 512]               0
               up-59          [-1, 8, 528, 512]               0
         Upsample-60        [-1, 8, 1056, 1024]               0
           Conv2d-61        [-1, 8, 1056, 1024]             136
      BatchNorm2d-62        [-1, 8, 1056, 1024]              16
      BatchNorm2d-63       [-1, 16, 1056, 1024]              32
             ReLU-64       [-1, 16, 1056, 1024]               0
           Conv2d-65        [-1, 8, 1056, 1024]           1,160
      BatchNorm2d-66        [-1, 8, 1056, 1024]              16
             ReLU-67        [-1, 8, 1056, 1024]               0
           Conv2d-68        [-1, 8, 1056, 1024]             584
   residual_block-69        [-1, 8, 1056, 1024]               0
               up-70        [-1, 8, 1056, 1024]               0
           Conv2d-71        [-1, 1, 1056, 1024]               9
          outconv-72        [-1, 1, 1056, 1024]               0
================================================================
Total params: 56,913
Trainable params: 56,913
Non-trainable params: 0
----------------------------------------------------------------
Input size (MB): 12.38
Forward/backward pass size (MB): 2442.00
Params size (MB): 0.22
Estimated Total Size (MB): 2454.59
----------------------------------------------------------------
\end{lstlisting}
and for plane 1:
\begin{lstlisting}
    ----------------------------------------------------------------
        Layer (type)               Output Shape         Param #
================================================================
            Conv2d-1        [-1, 8, 1400, 1024]              32
       BatchNorm2d-2        [-1, 8, 1400, 1024]              16
            Conv2d-3        [-1, 8, 1400, 1024]             224
       BatchNorm2d-4        [-1, 8, 1400, 1024]              16
              ReLU-5        [-1, 8, 1400, 1024]               0
            Conv2d-6        [-1, 8, 1400, 1024]             584
            inconv-7        [-1, 8, 1400, 1024]               0
            Conv2d-8         [-1, 16, 700, 512]             144
       BatchNorm2d-9         [-1, 16, 700, 512]              32
      BatchNorm2d-10        [-1, 8, 1400, 1024]              16
             ReLU-11        [-1, 8, 1400, 1024]               0
           Conv2d-12         [-1, 16, 700, 512]           1,168
      BatchNorm2d-13         [-1, 16, 700, 512]              32
             ReLU-14         [-1, 16, 700, 512]               0
           Conv2d-15         [-1, 16, 700, 512]           2,320
   residual_block-16         [-1, 16, 700, 512]               0
             down-17         [-1, 16, 700, 512]               0
           Conv2d-18         [-1, 32, 350, 256]             544
      BatchNorm2d-19         [-1, 32, 350, 256]              64
      BatchNorm2d-20         [-1, 16, 700, 512]              32
             ReLU-21         [-1, 16, 700, 512]               0
           Conv2d-22         [-1, 32, 350, 256]           4,640
      BatchNorm2d-23         [-1, 32, 350, 256]              64
             ReLU-24         [-1, 32, 350, 256]               0
           Conv2d-25         [-1, 32, 350, 256]           9,248
   residual_block-26         [-1, 32, 350, 256]               0
             down-27         [-1, 32, 350, 256]               0
           Conv2d-28         [-1, 32, 175, 128]           1,056
      BatchNorm2d-29         [-1, 32, 175, 128]              64
      BatchNorm2d-30         [-1, 32, 350, 256]              64
             ReLU-31         [-1, 32, 350, 256]               0
           Conv2d-32         [-1, 32, 175, 128]           9,248
      BatchNorm2d-33         [-1, 32, 175, 128]              64
             ReLU-34         [-1, 32, 175, 128]               0
           Conv2d-35         [-1, 32, 175, 128]           9,248
   residual_block-36         [-1, 32, 175, 128]               0
             down-37         [-1, 32, 175, 128]               0
         Upsample-38         [-1, 32, 350, 256]               0
           Conv2d-39         [-1, 16, 350, 256]           1,040
      BatchNorm2d-40         [-1, 16, 350, 256]              32
      BatchNorm2d-41         [-1, 64, 350, 256]             128
             ReLU-42         [-1, 64, 350, 256]               0
           Conv2d-43         [-1, 16, 350, 256]           9,232
      BatchNorm2d-44         [-1, 16, 350, 256]              32
             ReLU-45         [-1, 16, 350, 256]               0
           Conv2d-46         [-1, 16, 350, 256]           2,320
   residual_block-47         [-1, 16, 350, 256]               0
               up-48         [-1, 16, 350, 256]               0
         Upsample-49         [-1, 16, 700, 512]               0
           Conv2d-50          [-1, 8, 700, 512]             264
      BatchNorm2d-51          [-1, 8, 700, 512]              16
      BatchNorm2d-52         [-1, 32, 700, 512]              64
             ReLU-53         [-1, 32, 700, 512]               0
           Conv2d-54          [-1, 8, 700, 512]           2,312
      BatchNorm2d-55          [-1, 8, 700, 512]              16
             ReLU-56          [-1, 8, 700, 512]               0
           Conv2d-57          [-1, 8, 700, 512]             584
   residual_block-58          [-1, 8, 700, 512]               0
               up-59          [-1, 8, 700, 512]               0
         Upsample-60        [-1, 8, 1400, 1024]               0
           Conv2d-61        [-1, 8, 1400, 1024]             136
      BatchNorm2d-62        [-1, 8, 1400, 1024]              16
      BatchNorm2d-63       [-1, 16, 1400, 1024]              32
             ReLU-64       [-1, 16, 1400, 1024]               0
           Conv2d-65        [-1, 8, 1400, 1024]           1,160
      BatchNorm2d-66        [-1, 8, 1400, 1024]              16
             ReLU-67        [-1, 8, 1400, 1024]               0
           Conv2d-68        [-1, 8, 1400, 1024]             584
   residual_block-69        [-1, 8, 1400, 1024]               0
               up-70        [-1, 8, 1400, 1024]               0
           Conv2d-71        [-1, 1, 1400, 1024]               9
          outconv-72        [-1, 1, 1400, 1024]               0
================================================================
Total params: 56,913
Trainable params: 56,913
Non-trainable params: 0
----------------------------------------------------------------
Input size (MB): 16.41
Forward/backward pass size (MB): 3237.50
Params size (MB): 0.22
Estimated Total Size (MB): 3254.12
----------------------------------------------------------------
\end{lstlisting}
The best performing network for SBND is also a U-ResNet, summary for both planes:
\begin{lstlisting}
----------------------------------------------------------------
        Layer (type)               Output Shape         Param #
================================================================
            Conv2d-1         [-1, 32, 992, 857]             128
       BatchNorm2d-2         [-1, 32, 992, 857]              64
            Conv2d-3         [-1, 32, 992, 857]             896
       BatchNorm2d-4         [-1, 32, 992, 857]              64
              ReLU-5         [-1, 32, 992, 857]               0
            Conv2d-6         [-1, 32, 992, 857]           9,248
            inconv-7         [-1, 32, 992, 857]               0
            Conv2d-8         [-1, 64, 496, 429]           2,112
       BatchNorm2d-9         [-1, 64, 496, 429]             128
      BatchNorm2d-10         [-1, 32, 992, 857]              64
             ReLU-11         [-1, 32, 992, 857]               0
           Conv2d-12         [-1, 64, 496, 429]          18,496
      BatchNorm2d-13         [-1, 64, 496, 429]             128
             ReLU-14         [-1, 64, 496, 429]               0
           Conv2d-15         [-1, 64, 496, 429]          36,928
   residual_block-16         [-1, 64, 496, 429]               0
             down-17         [-1, 64, 496, 429]               0
           Conv2d-18        [-1, 128, 248, 215]           8,320
      BatchNorm2d-19        [-1, 128, 248, 215]             256
      BatchNorm2d-20         [-1, 64, 496, 429]             128
             ReLU-21         [-1, 64, 496, 429]               0
           Conv2d-22        [-1, 128, 248, 215]          73,856
      BatchNorm2d-23        [-1, 128, 248, 215]             256
             ReLU-24        [-1, 128, 248, 215]               0
           Conv2d-25        [-1, 128, 248, 215]         147,584
   residual_block-26        [-1, 128, 248, 215]               0
             down-27        [-1, 128, 248, 215]               0
           Conv2d-28        [-1, 128, 124, 108]          16,512
      BatchNorm2d-29        [-1, 128, 124, 108]             256
      BatchNorm2d-30        [-1, 128, 248, 215]             256
             ReLU-31        [-1, 128, 248, 215]               0
           Conv2d-32        [-1, 128, 124, 108]         147,584
      BatchNorm2d-33        [-1, 128, 124, 108]             256
             ReLU-34        [-1, 128, 124, 108]               0
           Conv2d-35        [-1, 128, 124, 108]         147,584
   residual_block-36        [-1, 128, 124, 108]               0
             down-37        [-1, 128, 124, 108]               0
         Upsample-38        [-1, 128, 248, 216]               0
           Conv2d-39         [-1, 64, 248, 215]          16,448
      BatchNorm2d-40         [-1, 64, 248, 215]             128
      BatchNorm2d-41        [-1, 256, 248, 215]             512
             ReLU-42        [-1, 256, 248, 215]               0
           Conv2d-43         [-1, 64, 248, 215]         147,520
      BatchNorm2d-44         [-1, 64, 248, 215]             128
             ReLU-45         [-1, 64, 248, 215]               0
           Conv2d-46         [-1, 64, 248, 215]          36,928
   residual_block-47         [-1, 64, 248, 215]               0
               up-48         [-1, 64, 248, 215]               0
         Upsample-49         [-1, 64, 496, 430]               0
           Conv2d-50         [-1, 32, 496, 429]           4,128
      BatchNorm2d-51         [-1, 32, 496, 429]              64
      BatchNorm2d-52        [-1, 128, 496, 429]             256
             ReLU-53        [-1, 128, 496, 429]               0
           Conv2d-54         [-1, 32, 496, 429]          36,896
      BatchNorm2d-55         [-1, 32, 496, 429]              64
             ReLU-56         [-1, 32, 496, 429]               0
           Conv2d-57         [-1, 32, 496, 429]           9,248
   residual_block-58         [-1, 32, 496, 429]               0
               up-59         [-1, 32, 496, 429]               0
         Upsample-60         [-1, 32, 992, 858]               0
           Conv2d-61         [-1, 32, 992, 857]           2,080
      BatchNorm2d-62         [-1, 32, 992, 857]              64
      BatchNorm2d-63         [-1, 64, 992, 857]             128
             ReLU-64         [-1, 64, 992, 857]               0
           Conv2d-65         [-1, 32, 992, 857]          18,464
      BatchNorm2d-66         [-1, 32, 992, 857]              64
             ReLU-67         [-1, 32, 992, 857]               0
           Conv2d-68         [-1, 32, 992, 857]           9,248
   residual_block-69         [-1, 32, 992, 857]               0
               up-70         [-1, 32, 992, 857]               0
           Conv2d-71          [-1, 1, 992, 857]              33
          outconv-72          [-1, 1, 992, 857]               0
================================================================
Total params: 893,505
Trainable params: 893,505
Non-trainable params: 0
----------------------------------------------------------------
Input size (MB): 9.73
Forward/backward pass size (MB): 7647.93
Params size (MB): 3.41
Estimated Total Size (MB): 7661.07
----------------------------------------------------------------
\end{lstlisting}
    